\documentclass[12pt,a4paper]{article}
\usepackage{indentfirst}
\usepackage{mathrsfs}
\usepackage{amsbsy}
\usepackage{amsmath,bm}
\usepackage{amssymb}
\usepackage{multicol}
\usepackage{times,epsfig,verbatim,lscape}
\usepackage{amsthm}
\usepackage{graphicx}
\usepackage{float}
\usepackage{subfigure}
\usepackage{color}
\usepackage{multirow}
\usepackage[textwidth=14.5cm]{geometry}
\usepackage{blindtext}
\usepackage{blkarray}
\usepackage{booktabs}%表格
\usepackage[titletoc]{appendix}
\usepackage[round]{natbib}
\usepackage{epstopdf}
\usepackage[tiny]{titlesec}%
\usepackage{titlesec}
       \titleformat{\section}
             {\normalfont\large\bfseries}{\thesection}{11pt}{\large}
\usepackage{titlesec}
      \titleformat{\subsection}
            {\normalfont\bfseries}{\thesubsection}{11pt}
\allowdisplaybreaks
\renewcommand{\baselinestretch} {1.2}
\makeatletter 

\def\singlespace{\def\baselinestretch{1}\@normalsize}

\def\non{\nonumber}
\def\bse{\begin{eqnarray*}}
\def\ese{\end{eqnarray*}}
\def\be{\begin{eqnarray}}
\def\ee{\end{eqnarray}}
\def\bsq{\begin{equation*}}
\def\esq{\end{equation*}}
\def\bq{\begin{equation}}
\def\eq{\end{equation}}

\numberwithin{equation}{section}
\renewcommand{\theequation} {\arabic{section}.\arabic{equation}}

\renewcommand{\hat}{\widehat}

\newtheorem{condition}{\underline{\bf Assumption}}

\newtheorem{remark}{\underline{\bf Remark}}
\allowdisplaybreaks
\def\tit.arg{\textbf{An outlier-robust model averaging approach by Mallows-type criterion }}
\DeclareMathOperator*{\argmin}{argmin}
\def\shorttitle.arg{
JMA for Composite Quantile Regression
}

\def\key.arg{Asymptotic optimality; Mallows's $C_{p}$; Model averaging; Robust prediction.
}

\def\abst.arg{
Model averaging is an alternative to model selection for dealing with model uncertainty, which is widely used and very valuable. However, most of the existing model averaging methods are proposed based on the least squares loss function, which could be very sensitive to the presence of outliers in the data.
In this paper, we propose an outlier-robust model averaging approach by Mallows-type criterion. The key idea is to develop weight choice criteria by minimising an estimator of the expected prediction error for the function being convex with an unique minimum, and twice differentiable in expectation, rather than the
expected squared error. The robust loss functions, such as least absolute deviation and Huber's function, reduce the effects of large residuals and poor samples.
Simulation study and real data analysis are conducted to demonstrate the finite-sample performance of our estimators and compare them with other model selection and averaging methods.
}

\def\author.arg{Miaomiao Wang$^{a,b,c,}$\footnote{Email: wangmm@amss.ac.cn}, Guohua Zou$^{d}$
\vskip 0.2cm
{\it \small $^{a}$School of Chinese Pharmacy, Beijing University of Chinese Medicine, Beijing, China. \\
$^{b}$Academy of Mathematics and Systems Science, Chinese Academy of Sciences, Beijing, China. \\
$^{c}$University of the Chinese Academy of Sciences, Beijing, China. \\
$^{d}$School of Mathematical Sciences, Capital Normal University, Beijing, China.
}\\
\medskip
}

\begin{document}

\thispagestyle{empty}

\begin{center}
{\large \tit.arg}%\footnote{Corresponding author: {Xinyu Zhang (xinyu@amss.ac.cn)}.}

\vskip 3mm

\author.arg
\end{center}
%\centerline{\today}
\vskip 3mm \centerline{\small ABSTRACT} \abst.arg
\\


\clearpage\pagebreak\newpage
\pagenumbering{arabic}
\setcounter{page}{1}

\section{Introduction}\label{sec1}
\setcounter{equation}{0}
\baselineskip=24pt
%第一段：异常值模型选择问题，文献，再就是还有很多对异常值稳健的Mallows type 模型选择方法。
Any applied statistician who has analysed lots of sets of actual data may encounter ``outliers". An outlier is an observation point that deviates significantly from most of the observed values, so that we suspect that it arises from a different mechanism. For more discussion on outliers, please refer to \cite{Hawkins1980}.
Outliers in the sample can have a large impact on some common statistical methods.
For example, outliers can affect the results of the least squares method, resulting in significant deviations in the sample mean. It has a serious impact on model selection and prediction.
Therefore, the study of the problem of outliers has increasingly attracted the attention of statisticians.

Outliers robust model selection is an important research direction of robust statistics, and there are a substantial literature on this subject.
These methods are roughly divided into two categories.
On one hand, some approaches are based on resampling methods,
such as cross-validation or the bootstrap.
For example, extending the work of \cite{Shao1993Linear}, \cite{Ronchetti1997Robust} developed a robust model selection technique for regression based on cross validation (\cite{Stone1974Cross}).
%\cite{Qian1998On} derived a criterion based on generalized Huberization
%and on the newly developed theory of stochastic complexity. Under general conditions, they studied the asymptotic properties concerning strong consistency of selecting the optimal model of the criterion.
Besides, \cite{Wisnowski2003Resampling} proposed a variable selection method for robust regression by combining robust estimation and resampling variable selection techniques.

On the other hand, most statisticians modified the popular criteria.
For example, \cite{Hampel1983Some} and \cite{Ronchetti1985Robust} suggested the robust version of Akaike Information Criterion (AIC, \cite{Akaike1973Information}) model selection procedure and investigated the properties of it.
\cite{Ronchetti1994A} presented a modified version of Mallows's $C_{p}$ (\cite{Mallows1973Some}) by weighted residual sum of squares. It allows us choose a model that fits most of the data by considering the existence of outliers.
\cite{Ronchetti1997Robustness} reviewed this criterion as well as some other approaches. He stressed that there remain much work to be done, such as robust model
selection in time series and developing other robust model selection procedures.
Following \cite{Ronchetti1994A}, \cite{Sommer1995Robust} implemented the robust version of Mallows's $C_{p}$ for Mallows-type estimators, which has a weight function that affects the function of the position and the function of the residual.
\cite{Sommer1996Variables} presented a new variables selection criterion based on the Wald test statistic (\cite{Wald1944On}).
\cite{Agostinelli2002Robust}, using weighted likelihood methodology developed by \cite{Agostinelli1998A}, introduced the robust model selection procedures by
modification of the AIC and Mallows's $C_{p}$.
\cite{M2005Outlier} proposed a new robust method based on combining a robust penalized criterion, that is very much like Bayesian information criterion (BIC, \cite{Schwarz1978The}), and a robust conditional expected prediction loss function that is estimated using a stratified bootstrap.

However, no matter which selection method is used, the search for the best model will identify the existence of multiple candidate models, which means that the level of uncertainty associated with model selection is usually ignored when reporting accurate estimators. An overly complex model may make the variance of the estimation or prediction too large, while an excessive simple model may lead to bias in the estimation or prediction. One way to solve the uncertainty of model selection is frequentist model averaging (FMA, \cite{hjort2003averaging}), where the estimation of unknown parameters is based on a set of weighted models rather than a single model. The main research questions are to find the model weight criterion and the statistical inference of the model average estimators.
Over the past two decades, there has been developed a substantial amount of FMA weight selection algorithms, including weighting by modification of popular model selection criteria \citep{Buckland1997Model, claeskens2006logit,  hjort2006focused, zhang2011focused, zhang2012focused}, adaptive regression by mixing (ARM, \citep{yang2001adaptive}) , the Mallows criterion \citep{hansen2007least, hansen2008joe, wan2010least}, MSE minimization \citep{liang2011optimal, wan2014ijf, zhang2014mixed}, cross-validation (CV) or jackknife procedures \citep{hansen2012jackknife, zhang2013jackknife, ando2014model, lu2015jackknife}, and minimization of Kullback-Leibler type measures \citep{zhang2015kullback, zhang2016optimal}. Among these criteria, the modification of Mallows criterion is developed early and most widely used, applied to linear regression model,  linear mixed-effects models and so on. However, since these methods are constructed based on the least squares loss function, which could be very sensitive to the presence of outliers
in the data.
%第二段：同样虽然稳健，但是同样像一般的模型选择方法一样忽略了模型选择的不确定性，很多缺点，列举。\\
%第三段：模型平均解决模型选择不确定性的问题，其中MMA方法应用广泛 ，有效，列举文献，但都是基于平方损失函数，对异常值并不稳健。Liu 提出的只是对异方差稳健的方法。\\

The need for robust averaging procedures is obvious because one cannot estimate the parameters robustly and apply unmodified classical model averaging procedures.
In this paper, we propose an outlier-robust model averaging approach by Mallows-type criterion. The key idea is to estimate model averaging weight by minimising an estimator of the expected prediction error for the function being convex with a unique minimum, and twice differentiable in expectation, rather than the expected squared error. The robust loss functions, such as least absolute deviation and Huber's function, reduce the effects of large residuals and poor samples.
%第四段：我们提出对异常值稳健的Mallows type MA 方法，谈谈想法，变化损失函数使其稳健。两种证明方法，表现等等。\\

The reminder of the article is organized as follows. Section \ref{sec:MFMAE} describes the model framework. In Section \ref{sec:Weight}, we present weight selection criteria. Section \ref{sec:methods} reviews three robust version of Mallows's $C_{p}$ and Mallows model average method for comparison.
Section \ref{sec:simu} investigates the finite sample performance of our proposed method, and then we apply the proposed method to three data examples in Section \ref{sec:data}. Some concluding remarks are contained in Section \ref{sec:conc}. Derivation of technical results are given in the Appendix.

\section{Model Framework and Model Average Estimator }\label{sec:MFMAE}
Suppose random sample $y_{i}$ is generated by a linear regression
\be
y_{i}=\mu_{i}+\varepsilon_{i},~~~~i=1,\ldots,n,\non\ee
where $\mu_{i}=\boldsymbol{x}_{i}^{T}\Theta$, $\boldsymbol{x}_{i}=(x_{i1},\ldots,x_{ip})^{T}$ is $p$-dimensional vector of covariates
 and $\Theta=(\theta_{1},\ldots,\theta_{p})^{T}$ is the corresponding coefficient vector;
$\varepsilon_{i},i=1,\ldots,n$ are independent observations with mean $0$, variance $\sigma^{2}$ and
density $f(\cdot)$ corresponding to the unknown cumulative distribution function $F(\cdot)$;
$\varepsilon_{i}$ is independent to $\boldsymbol{x}_{i}$.
%Consider the problem of estimating $\mu$ by a linear model of the
%form $\boldsymbol{x}^{T}\Theta$ with $\Theta\in \mathbb{R}^{p}$.

Suppose the loss function is $\rho(e)$. We place the same assumptions as in \cite{Burman1995A} on it.% and the error distribution as follows.
\begin{condition}\label{con:rho}
$\rho(e)$ is convex with unique minimum at $0$.
\end{condition}
\begin{condition}\label{con:rho1}
$\mathrm{E}\rho_{1}(\varepsilon)=0$, where $\rho_{1}$ is the derivative of $\rho$.
\end{condition}

Write the $m^{th}$ candidate model as
 \begin{eqnarray}
y_{i}=\boldsymbol{x}_{i(m)}^{T}\Theta_{(m)}+\varepsilon_{i}=\sum_{j=1}^{k_{m}}{\theta_{j(m)}x_{ij(m)}+\varepsilon_{i(m)}},
\end{eqnarray}
where $k_{m}$ is the number of covariates, $\Theta_{(m)}=(\theta_{1(m)},\ldots,\theta_{k_{m}(m)})^{T}$ and $\boldsymbol{x}_{i(m)}=(x_{i1(m)},\ldots,x_{ik_{m}(m)})^{T}$ with $x_{ij(m)}$ being a covariate and $\theta_{j(m)}$ beging the corresponding coefficient, $j=1,\ldots,k_{m}$.  The estimator of $\Theta_{(m)}$ in the above model is
\begin{eqnarray}
\widehat{\Theta}_{(m)}= \operatorname*{\argmin\limits}_{\Theta_{(m)}\in \mathbb{R}^{k_{m}}}\sum_{i=1}^{n}\rho(y_{i}-\boldsymbol{x}_{i(m)}^{T}\Theta_{(m)}).
\end{eqnarray}
Suppose the number of the candidate models is $M$.
 Let $\widehat{\varepsilon}_{i(m)}= y_{i}-\boldsymbol{x}_{i(m)}^{T}\widehat{\Theta}_{(m)}$, $\boldsymbol{w}= (w_{1},\ldots,w_{M})^{T}$ be a  weight vector in the unit simplex of $\mathbb{R}^{M}$ and $\mathcal{W}=\{\boldsymbol{w}\in[0,1]^{M}:0\leq w_{k}\leq 1, \sum_{m=1}^{M}w_{m}=1\}$. The model averaging estimator of $\mu_{i}$ is thus
\be \label{eq:avg}
\widehat{\mu}_{i}(\boldsymbol{w})=\sum_{m=1}^{M}w_{m}\boldsymbol{x}_{i(m)}^{T}\widehat{\Theta}_{(m)},
\ee
where the weight $\boldsymbol{w}$ is unknown.
Further, in next section, we will develop a Mallows-type weight estimator, denoted by $\widehat{\boldsymbol{w}}$, is obtained by minimizing some Mallows-type weight selection criteria.
Substituting $\widehat{\boldsymbol{w}}$ for $\boldsymbol{w}$ in (\ref{eq:avg}) results in the following Mallows-type model average (MTMA) estimator of $\mu_{i}$:
\be
\widehat{\mu}_{i}\left(\widehat{\boldsymbol{w}}\right)=\sum_{m=1}^{M}\widehat{w}_{m}\boldsymbol{x}_{i(m)}^{T}\widehat{\Theta}_{(m)}.
\ee
\section{Weight Choice Schemes}\label{sec:Weight}
In this section, we will present weight selection criteria in two cases. One is the case where the design matrix is fixed, and the other is the case where the design matrix is random. The derivation of these two criteria are in \ref{sec:A} and \ref{sec:B}. Before that, let us give another hypothesis and some notations.
Let $\mathrm{E}_{\varepsilon}$ be the expectation taken on random variable $\varepsilon$.
\begin{condition}\label{con:R2}
$R(t): =\mathrm{E}_{\varepsilon}\rho(\varepsilon+t)$ is twice differentiable with second derivative $R_{2}$.
\end{condition}
Define $
\Theta^{*}=\argmin_{\Theta\in\mathbb{R}^{p}}\sum_{i=1}^{n}\mathrm{E}\left\{\rho\left(y_{i}-\boldsymbol{x}_{i}^{T}\Theta\right)\right\}
$,
$
\widehat{\Theta}=\argmin_{\Theta\in\mathbb{R}^{p}}\sum_{i=1}^{n}\rho\left(y_{i}-\boldsymbol{x}_{i}^{T}\Theta\right)
$,
$\Theta_{(m)}^{*}=\argmin_{\Theta_{(m)}\in\mathbb{R}^{k_{m}}}\sum_{i=1}^{n}\mathrm{E}\left\{\rho\left(y_{i}-\boldsymbol{x}_{i(m)}^{T}\Theta_{(m)}\right)\right\}$
and $\hat{\varepsilon}_{i}(\boldsymbol{w})=y_{i}-\widehat{\mu}_{i}(\boldsymbol{w})$.

\subsection{Weight selection criteria with fixed design matrix }
Assume that the design matrix $\boldsymbol{X}=\left(\boldsymbol{x}_{1}^{T},\ldots,\boldsymbol{x}_{n}^{T}\right)^{T}$ is fixed and known. We generate the out-of-sample observations $\{\widetilde{y}_{i}\}_{i=1}^{n}$ and then
evaluate the final prediction error measure. We propose a general Mallows-type criterion (MTC) for the model average estimator that is constructed as
\be\label{C:fixed}
C_{n}(\boldsymbol{w})=\sum_{i=1}^{n}\rho\left(\hat{\varepsilon}_{i}(\boldsymbol{w})\right)
+C_{\rho}\sum_{m=1}^{M}w_{m}k_{m},
\ee
where
\be
C_{\rho}=\frac{\sum_{i=1}^{n}\mathrm{var}\left\{\rho_{1}\left(y_{i}-\boldsymbol{x}_{i}^{T}\Theta^{*}\right)\right\}}
{\sum_{i=1}^{n}R_{2}\left(\mu_{i}-\boldsymbol{x}_{i}^{T}\Theta^{*}\right)}
\ee
and $\Theta^{*}$ can be estimated by $\widehat{\Theta}$. Then $C_{\rho}$ can be replaced by sample estimators under different loss functions. We provide the derivation of criterion (\ref{C:fixed}) in \ref{sec:A}. The general Mallows-type criterion is a sum of $\rho$-residuals of the model averaging estimator plus a term that brings in weighted the dimension of the fitted models.

The Mallows-type weight vector $\widehat{\boldsymbol{w}}=\left(\widehat{w}_{1},\ldots,\widehat{w}_{M}\right)^{T}$ is obtained by choosing $\boldsymbol{w}\in\mathcal{W}$ such that
\be
\widehat{\boldsymbol{w}}=\operatorname*{\argmin\limits}_{\boldsymbol{w}\in\mathcal{W}}C_{n}(\boldsymbol{w}).
\ee
Substituting $\widehat{\boldsymbol{w}}$ for $\boldsymbol{w}$ in (\ref{eq:avg}) results in the following MTMA estimator of $\mu_{i}$:
\be
\widehat{\mu}_{i}\left(\widehat{\boldsymbol{w}}\right)=\sum_{m=1}^{M}\widehat{w}_{m}\boldsymbol{x}_{i(m)}^{T}\widehat{\Theta}_{(m)}.
\ee
\begin{remark}
Our criterion degenerates into the criterion in \emph{\cite{Burman1995A}} when one component of the vector $\boldsymbol{w}$ is equal to one and the remaining weights are equal to zero.
\end{remark}
Note that the factor $C_{\rho}$ is unknown and is similar as that in \cite{Burman1995A}. Following \cite{Burman1995A}, we will present approximate estimators of $C_{\rho}$ for several commonly used loss functions
and it will be used in simulation study and real data analysis.
\subsubsection{Examples}
\noindent\underline{Square loss:}
When $\rho(t)=t^{2} $, $\rho_{1}(t)=2t$ and $R_{2}=2$. Noting that $R_{2}$ does not involve model bias,
therefore, when the errors have mean $0$ and variance $\sigma^{2}$, we have $C_{\rho}=2\sigma^{2}$ and
\be
C_{n}(\boldsymbol{w})=\sum_{i=1}^{n}\left\{\hat{\varepsilon}_{i}(\boldsymbol{w})\right\}^{2}
+2\sigma^{2}\sum_{m=1}^{M}w_{m}k_{m}.
\ee
We find that the general criterion put forth here coincides
with the Mallows model average criterion proposed by \cite{hansen2007least} and
\cite{wan2010least}. Suppose the $M^{th}$ model is the largest model.
In practice, $\sigma^{2}$ is unknown and can be estimated by
$\widehat{\sigma}^{2}=\sum_{i=1}^{n}\left(y_{i}-\textbf{x}_{i(M)}^{T}\Theta_{(M)}\right)^{2}\big/(n-k_{M})$.

\noindent\underline{Absolute loss:}
When the squared loss function is replaced with the absolute loss function, the first derivative of $\rho(t)$ is
\be
\rho_{1}(t)=
\begin{cases}
~1& t\geq0,\\
-1& t<0.
\end{cases}
\ee
We find $\rho(t)$ is no longer differentiable. It is differentiable almost everywhere and that the expected loss is twice differentiable so that it meets the assumptions.
Hence,
\be
&&\mathrm{Var}\left\{\rho_{1}\left(y_{i}-\boldsymbol{x}_{i}^{T}\Theta^{*}\right)\right\}\non\\
&=&1-\left\{P\left(\varepsilon_{i}\geq\boldsymbol{x}_{i}^{T}\Theta^{*}-\mu_{i}\right)
-P\left(\varepsilon_{i}<\boldsymbol{x}_{i}^{T}\Theta^{*}-\mu_{i}\right)\right\}^{2}\non\\
&=&4F\left(\boldsymbol{x}_{i}^{T}\Theta^{*}-\mu_{i}\right)\left\{1-F\left(\boldsymbol{x}_{i}^{T}\Theta^{*}-\mu_{i}\right)\right\}.
\ee
Provided that $F$ has median $0$ and continuous density $f$, then $R_{2}(t)=2f(t)$.
Therefore
\be
C_{\rho}=\frac{\sum_{i=1}^{n}4F\left(\boldsymbol{x}_{i}^{T}\Theta^{*}-\mu_{i}\right)\left\{1-F\left(\boldsymbol{x}_{i}^{T}\Theta^{*}-\mu_{i}\right)\right\}}
{\sum_{i=1}^{n}2f\left(\mu_{i}-\boldsymbol{x}_{i}^{T}\Theta^{*}\right)}.
\ee
This expression simplifies greatly if the model bias is $0$, for then
$
\widehat{C}_{\rho}=1/2f(0),
$
which is an approximation for $C_{\rho}$.
In the simulation study and real data analysis, the density $f(0)$ is estimated based on the Epanechnikov kernel with bandwidth the semi-interquartile range of
the residuals. We label the above method as $\text{MA}_{A}^{c}$.

\noindent\underline{Huber's function:}
The Huber's function is
\be
\rho(t)=
\begin{cases}
t^{2}& |t|\leq c,\\
2c|t|-c^{2}& |t|> c
\end{cases}
\ee
with a constant $c$, that is proposed by \cite{Huber1964Robust} in robust regression and is smooth yet linear in the tails.
Let $\boldsymbol{1}(\cdot)$ denote the indicator of event $\cdot$. As shown in \cite{Burman1995A},
the Huber's function meets the Assumptions \ref{con:rho} and \ref{con:rho1} if
$\mathrm{E}\left\{\varepsilon \boldsymbol{1}(|\varepsilon|\leq c)\right\}=0$
and
$F(-c)=1-F(c)$. Any distribution, that is symmetric about the origin, meets these conditions.
Note that  $R_{2}(t)=2P(|\varepsilon+t|\leq c)$
and $$\mathrm{Var}\left\{\rho_{1}\left(\varepsilon\right)\right\}
=4\mathrm{Var}\left\{\varepsilon\boldsymbol{1}(|\varepsilon|\leq c)\right\}
+4c^{2}P(|\varepsilon+t|> c).$$
Therefore, \cite{Burman1995A} suggested the estimator for $C_{\rho}$ as follows
\be
\widehat{C}_{\rho}=\frac{2\sum_{i=1}^{n}\widehat{\varepsilon}_{i}^{2}\boldsymbol{1}(|\widehat{\varepsilon}_{i}|\leq c)+2c^{2}\sum_{i=1}^{n}\boldsymbol{1}(|\widehat{\varepsilon}_{i}|> c)}
{\sum_{i=1}^{n}\boldsymbol{1}(|\widehat{\varepsilon}_{i}|\leq c)}.
\ee
Let $\widehat{\varepsilon}_{i}=\widehat{\varepsilon}_{i(M)}$ and $c=1.345$ in the simulation study and real data analysis. We label this method as $\text{MA}_{H}^{c}$.
\subsection{Weight selection criteria with random design matrix}
Given $\boldsymbol{x}_{1},\ldots,\boldsymbol{x}_{n}$ believed independently and identically distributed, we generate the out-of-sample observations $\{\widetilde{\boldsymbol{x}}_{i},\widetilde{y}_{i}\}_{i=1}^{n}$ and they are the copies of $\{\boldsymbol{x}_{i},y_{i}\}_{i=1}^{n}$.
Then a general Mallows-type criterion for the model average estimator is
\be\label{C:random}
\widetilde{C}_{n}(\boldsymbol{w})=\sum_{i=1}^{n}\rho\left(\hat{\varepsilon}_{i}(\boldsymbol{w})\right)
+\sum_{m=1}^{M}w_{m}k_{m}C_{\rho(m)}
\ee
where
$$C_{\rho(m)}=\mathrm{E}\left\{\rho_{1}\left(y_{i}-\boldsymbol{x}_{i(m)}^{T}\Theta_{(m)}^{*}\right)\rho_{1}\left(y_{i}-\sum_{m=1}^{M}w_{m}\boldsymbol{x}_{i(m)}^{T}\Theta_{(m)}^{*}\right)
\left\{\frac{1}{n}\sum_{i=1}^{n}R_{2}\left(\mu_{i}-\boldsymbol{x}_{i(m)}^{T}\Theta_{(m)}^{*}\right)\right\}^{-1}\right\}.$$
We provide the derivation of criterion (\ref{C:random}) in \ref{sec:B}.

The term $C_{\rho(m)}$
can be estimated by
\be\label{eq:Crhom}
\widehat{C}_{\rho(m)}=\frac{1}{n}\sum_{i=1}^{n}\left\{\rho_{1}\left(\widehat{\varepsilon}_{i(m)}\right)\rho_{1}\left(\sum_{m=1}^{M}w_{m}\widehat{\varepsilon}_{i(m)}\right)
\left\{\frac{1}{n}\sum_{i=1}^{n}\widehat{R}_{2}\left(\widehat{\varepsilon}_{i(m)}\right)\right\}^{-1}\right\},\ee
where
$\widehat{R}_{2}$ is an approximation for $R_{2}$.

The Mallows-type weight vector $\widehat{\widetilde{\boldsymbol{w}}}=\left(\widehat{\widetilde{w}}_{1},\ldots,\widehat{\widetilde{w}}_{M}\right)^{T}$ is obtained by choosing $\boldsymbol{\widetilde{w}}\in\mathcal{W}$ such that
\be
\widehat{\widetilde{\boldsymbol{w}}}=\operatorname*{\argmin\limits}_{\boldsymbol{w}\in\mathcal{W}}\widetilde{C}_{n}(\boldsymbol{w}).
\ee
Substituting $\widehat{\widetilde{\boldsymbol{w}}}$ for $\boldsymbol{w}$ in \eqref{eq:avg} results in the following MTMA estimator of $\mu_{i}$:
\be
\widehat{\mu}_{i}\left(\widehat{\widetilde{\boldsymbol{w}}}\right)=\sum_{m=1}^{M}\widehat{\widetilde{w}}_{m}\boldsymbol{x}_{i(m)}^{T}\widehat{\Theta}_{(m)}.
\ee
Note that $C_{\rho(m)}$ is variable under different models while $C_{\rho}$ is the same value for all models.
As above section, we will present approximate estimators of $C_{\rho(m)}$ for several commonly used loss functions as follows.
\subsubsection{Examples}

\noindent\underline{\text{Square loss}:}
For the least squares case, note that
$\rho_{1}(t)=2t$ and $R_{2}=2$, therefore,
$C_{\rho(m)}
=2\mathrm{E}\left\{\left(y_{i}-\boldsymbol{x}_{i(m)}^{T}\Theta_{(m)}^{*}\right)\left(y_{i}-\sum_{m=1}^{M}w_{m}\boldsymbol{x}_{i(m)}^{T}\Theta_{(m)}^{*}\right)\right\}$.
%\begin{eqnarray}
%&&\mathrm{E}\left\{\frac{1}{n}\sum_{i=1}^{n}\left(\widetilde{y}_{i}-\sum_{m=1}^{M}w_{m}\widetilde{\boldsymbol{x}}_{i(m)}^{T}\widehat{\Theta}_{(m)}\right)^{2}\right\}\nonumber\\
%&=&\mathrm{E}\left\{\frac{1}{n}\sum_{i=1}^{n}\left(\mu_{i}+\widetilde{\varepsilon}_{i}-\sum_{m=1}^{M}w_{m}\widetilde{\boldsymbol{x}}_{i(m)}^{T}\widehat{\Theta}_{(m)}\right)^{2}\right\}\nonumber\\
%&=&\mathrm{E}\left\{\frac{1}{n}\sum_{i=1}^{n}\left(\mu_{i}-\sum_{m=1}^{M}w_{m}\widetilde{\boldsymbol{x}}_{i(m)}^{T}\widehat{\Theta}_{(m)}\right)^{2}
%+\frac{1}{n}\sum_{i=1}^{n}\widetilde{\varepsilon}_{i}^{2}
%+\frac{2}{n}\sum_{i=1}^{n}\widetilde{\varepsilon}_{i}\left(\mu_{i}-\sum_{m=1}^{M}w_{m}\widetilde{\boldsymbol{x}}_{i(m)}^{T}\widehat{\Theta}_{(m)}\right)\right\}\nonumber\\
%&=&\sigma^{2}+\mathrm{E}\left\{\frac{1}{n}\sum_{i=1}^{n}\left(\mu_{i}-\sum_{m=1}^{M}w_{m}\widetilde{\boldsymbol{x}}_{i(m)}^{T}\widehat{\Theta}_{(m)}\right)^{2}\right\}.
%\end{eqnarray}
%Thus, for least squared loss function, estimating the weights with minimizing the average squared error is equivalent to
%minimizing the out-of-sample final prediction error.
To simplify, we use $\boldsymbol{x}_{i}^{T}\Theta$ to approximate $\boldsymbol{x}_{i(m)}^{T}\Theta_{(m)}^{*}$ and then
$C_{\rho(m)}\approx2\sigma^{2}$.
We can adopt the following Mallows criterion proposed by \cite{hansen2007least} for the model average estimator. That is
\be\label{eqCn:LS}
C_{n}^{s}(\boldsymbol{w})=\sum_{i=1}^{n}\left\{\hat{\varepsilon}_{i}(\boldsymbol{w})\right\}^{2}
+2\sigma^{2}\sum_{m=1}^{M}w_{m}k_{m}.
\ee
Suppose the $M^{th}$ model is the largest model.
In practice, $\sigma^{2}$ is unknown and can be estimated by
$\widehat{\sigma}^{2}=\sum_{i=1}^{n}\left(y_{i}-\textbf{x}_{i(M)}^{T}\Theta_{(M)}\right)^{2}\big/(n-k_{M})$.
To sum up, when we know that there are no outliers in the sample, we advocate the Mallows model average method.

\noindent\underline{Absolute loss:}
When the squared error is replaced with the absolute error, $R_{2}(t)=f(t)$, and then
$\widehat{R}_{2}\left(\widehat{\varepsilon}_{i(m)}\right)=\widehat{f}\left(\widehat{\varepsilon}_{i(m)}\right)$.
Therefore,
\be
\widehat{\widetilde{C}}_{\rho(m)}=\frac{1}{n}\sum_{i=1}^{n}\left\{\rho_{1}\left(\widehat{\varepsilon}_{i(m)}\right)\rho_{1}\left(\sum_{m=1}^{M}w_{m}\widehat{\varepsilon}_{i(m)}\right)
\left\{\frac{1}{n}\sum_{i=1}^{n}\widehat{f}\left(\widehat{\varepsilon}_{i(m)}\right)\right\}^{-1}\right\}
\ee
with function
\be
\rho_{1}(t)=
\begin{cases}
~1& t\geq0,\\
-1& t<0.\non
\end{cases}
\ee
We label this method as $\text{MA}_{A}$.

\noindent\underline{Huber's function:}
Using the Huber's function,
$R_{2}(t)=2\mathrm{P}(|\varepsilon+t|\leq c)$. Then
\be
\widehat{\widetilde{C}}_{\rho(m)}=\frac{1}{n}\sum_{i=1}^{n}\left\{\rho_{1}\left(\widehat{\varepsilon}_{i(m)}\right)\rho_{1}\left(\sum_{m=1}^{M}w_{m}\widehat{\varepsilon}_{i(m)}\right)
\left\{\frac{1}{n}\sum_{i=1}^{n}\widehat{R}_{2}\left(\widehat{\varepsilon}_{i(m)}\right)\right\}^{-1}\right\},
\ee
where $$\widehat{R}_{2}\left(\widehat{\varepsilon}_{i(m)}\right)
=\frac{2}{n}\sum_{j=1}^{n}\boldsymbol{1}\left(|\widehat{\varepsilon}_{j(M)}+\widehat{\varepsilon}_{i(m)}|\leq c\right)$$
and function
\be
\rho_{1}(t)=
\begin{cases}
~~2t& |t|\geq c,\\
~~2c&  t>  c,\\
-2c&   t< -c.\non
\end{cases}
\ee
Similarly, we label this method as $\text{MA}_{H}$.

\section{Methods for comparison}\label{sec:methods}
In this section, we review three robust version of Mallows's $C_{p}$ and Mallows model average method for comparison in simulation study and real data analysis.

\noindent\underline{\textbf{Robust model selection method 1}:} An M estimator $\widetilde{\Theta}_{(m)}$ for model $m$ is the solution of the equation
\be
\sum_{i=1}^{n}\eta\left(\boldsymbol{x}_{i(m)},y_{i}-\boldsymbol{x}_{i(m)}^{T}\Theta_{(m)}\right)\boldsymbol{x}_{i(m)}=0
\ee
for some function $\eta(\boldsymbol{x},r)$.
Let weights $$\hat{\varpi }_{i}=\varpi\left(\boldsymbol{x}_{i(m)},y_{i}-\boldsymbol{x}_{i(m)}^{T}\widetilde{\Theta}_{(m)}\right)
=\frac{\eta\left(\boldsymbol{x}_{i(m)},y_{i}-\boldsymbol{x}_{i(m)}^{T}\Theta_{(m)}\right)}{y_{i}-\boldsymbol{x}_{i(m)}^{T}\widetilde{\Theta}_{(m)}}.$$
In \cite{Ronchetti1994A},
they defined a robust version of Mallows's $C_{p}$ based on M estimator for regression models as follows:
\be
RC_{p}(k_{m})=\frac{W_{m}}{\widetilde{\sigma}^{2}}-\left(U_{m}-V_{m}\right),
\ee
where $W_{m}=\sum_{i=1}^{n}\hat{\varpi }_{i}^{2}\left(y_{i}-\boldsymbol{x}_{i(m)}^{T}\widetilde{\Theta}_{(m)}\right)^{2}$
is the weighted residual sum of squares.
Let $\eta'$ and $\varpi'$ denote the derivative of $\eta(\boldsymbol{x},r)$ and $\varpi(\boldsymbol{x},r)$ with respect to its second argument, respectively.
Define $Q=\mathrm{E}\left\{\eta^{2}(\boldsymbol{x},\varepsilon)\boldsymbol{x}\boldsymbol{x}^{T}\right\}$,
$\|\eta\|^{2}=\sum_{1\leq i\leq n}\eta^{2}(\boldsymbol{x}_{i},\varepsilon_{i})$,
$N=\mathrm{E}(\eta^{2}\eta'\boldsymbol{x}\boldsymbol{x}^{T})$,
$L=\mathrm{E}\left\{\varpi'\varepsilon(\varpi'\varepsilon+4\varpi)\boldsymbol{x}\boldsymbol{x}^{T}\right\}
=\mathrm{E}\left[\left\{(\eta')^{2}+2\eta'\varpi-3\varpi^{2}\right\}\boldsymbol{x}\boldsymbol{x}^{T}\right]$
and $R=\mathrm{E}(\varpi^{2}\boldsymbol{x}\boldsymbol{x}^{T})$.
Then we can define constants
$$U_{m}-V_{m}=\mathrm{E}\|\eta\|^{2}-2\mathrm{tr}(NM^{-1})+\mathrm{tr}(LM^{-1}QM^{-1})$$
and $V_{m}=\mathrm{tr}(RM^{-1}QM^{-1})$.
Besides, $\widetilde{\sigma}^{2}$ is a robust and consistent estimator of $\sigma^{2}$ in the full model estimated by  $W_{M}/U_{M}$.

Consider the weighted least squares estimation.
We label this robust model selection method by HC$_p$, which is based on the robust estimators with a weight function of Huber's type ($\varpi\left(\boldsymbol{x}_{i(m)},r_{i}\right)=\eta(r_{i})/r_{i}$).
While the robust model selection method is labeled by MC$_p$ based on the robust estimators with a weight function of Mallows's type ($\varpi\left(\boldsymbol{x}_{i(m)},r_{i}\right)=v(\boldsymbol{x}_{i(m)})\eta(r_{i})/r_{i}$), where
the factor dependent on $r_{i}$ is bounded and the factor dependent on $\boldsymbol{x}_{i(m)}$ satisfies $|v(\boldsymbol{x}_{i(m)})|<K/\|\boldsymbol{x}_{i(m)}\|^{2}$ for some constant $K$.
More details refer to \cite[chap.6]{hampel1986robust}.
We prefer to choose the models with values of $RC_{p}(k_{m})$ close to $V_{m}$
or smaller than $V_{m}$.

\noindent\underline{\textbf{Robust model selection method 2}:}
A general Akaike-type criterion in \cite{Burman1995A} is
\be
\bar{L}_{n}(k_{m})=\sum_{i=1}^{n}\rho\left(\hat{\varepsilon}_{i(m)}\right)
+C_{\rho}k_{m},
\ee
where
\be
C_{\rho}=\frac{\sum_{i=1}^{n}\mathrm{var}\left\{\rho_{1}(y_{i}-\boldsymbol{x}_{i}^{T}\Theta^{*})\right\}}
{\sum_{i=1}^{n}R_{2}\left(\mu_{i}-\boldsymbol{x}_{i}^{T}\Theta^{*}\right)}
\ee
and can be estimated for a variety of examples with loss function $\rho$.

We prefer to choose the models with the minimum value of $\bar{L}_{n}(k_{m})$.
This selection method is labeled by MS$_{H}$ when we use the Huber's function and is labeled by MS$_{A}$ when we use the absolute error.

\noindent\underline{\textbf{Robust model selection method 3}:}
Under the $m^{th}$ regression model,
let $z_{i(m)}(\Theta_{(m)})=y_{i}-\boldsymbol{x}_{i(m)}^{T}\Theta_{(m)}$, the residuals for the parameter $\Theta_{(m)}$, and $\widehat{F}_{n,m}(t;\Theta_{(m)})=\sum_{i=1}^{n}\boldsymbol{1}\{z_{i(m)}(\Theta_{(m)})<t\}/n$, the empirical cumulative distribution.
Based on weighted likelihood methodology developed by \cite{Agostinelli1998A},
the weight function $\varphi\left\{z_{i(m)}(\Theta_{(m)});\sigma,\widehat{F}_{n,m}(\Theta_{(m)})\right\}$ for sample $i$, depending on the $m^{th}$ model and empirical cumulative distribution, is constructed as
$$\varphi\left\{z_{i(m)}(\Theta_{(m)});\sigma,\widehat{F}_{n,m}(\Theta_{(m)})\right\}
=\min\left\{1,\frac{\left[A\left[\delta\left\{z_{i(m)}(\Theta_{(m)});\sigma,\widehat{F}_{n,m}(\Theta_{(m)})\right\}\right]+1\right]^{+}}
{\delta\left\{z_{i(m)}(\Theta_{(m)});\sigma,\widehat{F}_{n,m}(\Theta_{(m)})\right\}+1}\right\},$$
where $[\cdot]^{+}$ indicates the positive part of $\cdot$.
$\delta\left\{z_{i(m)}(\Theta_{(m)});\sigma,\widehat{F}_{n,m}(\Theta_{(m)})\right\}$ is called the Pearson residual and it is defined as
$$\delta\left\{z_{i(m)}(\Theta_{(m)});\sigma,\widehat{F}_{n,m}(\Theta_{(m)})\right\}
=\frac{\int k(z_{i(m)}(\Theta_{(m)});t,h)\mathrm{d}\widehat{F}_{n,m}(t;\Theta_{(m)})}{\int k(z_{i(m)}(\Theta_{(m)});t,h)\mathrm{d}F(t)}-1,$$
where $k(z_{i(m)}(\Theta_{(m)});t,h)$ is a kernel function and we use the Gaussian kernel function with smoothing parameter $h$.%0.032\widehat{\sigma}^{2}
The function $A(\cdot)$ is a Residual Adjustment Function (RAF, \cite{lindsay1994}), where we use Hellinger Residual Adjustment Function and it is defined as $A(\delta)=2\{\sqrt{\delta+1}-1\}$.
Hence, the weighted likelihood estimator of the parameters $\Theta_{(m)}$ and $\sigma$ are $\widehat{\Theta}_{(m)}^{w}$ and $\widehat{\sigma}$, which are the solution of the estimating equations:
\be
\sum_{i=1}^{n}\varphi\left\{z_{i(m)}(\Theta_{(m)});\sigma,\widehat{F}_{n,m}(\Theta_{(m)})\right\}
u\left\{z_{i(m)}(\Theta_{(m)});\sigma\right\}=0
\ee
and
\be
\sum_{i=1}^{n}\varphi\left\{z_{i(m)}(\Theta_{(m)});\sigma,\widehat{F}_{n,m}(\Theta_{(m)})\right\}
u_{\sigma}\left\{z_{i(m)}(\Theta_{(m)});\sigma\right\}=0,
\ee
where $$u\left\{z_{i(m)}(\Theta_{(m)});\sigma\right\}=(\partial/\partial\Theta_{(m)})\log f\left\{z_{i(m)}(\Theta_{(m)})\right\}$$ and
$$u_{\sigma}\left\{z_{i(m)}(\Theta);\sigma\right\}=(\partial/\partial\sigma)\log f\left\{z_{i(m)}(\Theta_{(m)})\right\}.$$
 Then, in the Gaussian kernel function, let smoothing parameter $h$ be $0.032\widehat{\sigma}^{2}$ in the simulation study and real data analysis.

Applying the weighted likelihood methodology, for robust regression, \cite{Agostinelli2002Robust} provided a direct extension of the Mallows's $C_{p}$, that is the weighted Mallows's $C_{p}$
\be
WC_{p}(k_{m})&=&\sum_{i=1}^{n}\varphi\left\{z_{i(m)}\left(\Theta_{(m)}^{*}\right);\widehat{\sigma},\widehat{F}_{n,m}\left(\Theta_{(m)}^{*}\right)\right\}z_{i(m)}\left(\widehat{\Theta}_{(m)}^{w}\right)^{2}
/\widehat{\sigma}^{2}\non\\
&&-\sum_{i=1}^{n}\varphi\left\{z_{i(m)}\left(\Theta_{(m)}^{*}\right);\widehat{\sigma},\widehat{F}_{n,m}\left(\Theta_{(m)}^{*}\right)\right\}+2k_{m}.
\ee
He recommended setting $\Theta_{(m)}^{*}=\widehat{\Theta}_{(M)}^{w}$ which estimated under the largest model.
We prefer the model with the minimum value of $WC_{p}(k_{m})$, $m\in\{1,\cdots,M\}$.
This selection method is labeled by WC$_{p}$.

\noindent\underline{\textbf{Mallows's model average method}:}
The Mallows model average estimators proposed by \citep{hansen2007least, hansen2008joe, wan2010least}
is based on the squared loss function $\rho(t)=t^{2}$ and as result it is sensitive to outliers.
The Mallows model average criterion is
\be
C_{n}(\boldsymbol{w})=\sum_{i=1}^{n}\left\{\hat{\varepsilon}_{i}(\boldsymbol{w})\right\}^{2}
+2\sigma^{2}\sum_{m=1}^{M}w_{m}k_{m},
\ee
where $\sigma^{2}$ is estimated by
$\widehat{\sigma}^{2}=\sum_{i=1}^{n}\left(y_{i}-\textbf{x}_{i(M)}^{T}\Theta_{(M)}\right)^{2}\big/(n-k_{M})$.
This model averaging method is labeled by MMA.

We evaluate the performance of the above methods with respect to the following final absolute prediction error measure across $R$ replications:
 $$\mathrm{APE}=\frac{1}{R}\sum_{r=1}^{R}\mathrm{PE}(r),$$
where
$$\mathrm{PE}(r)=\frac{1}{n_s}\sum_{s=1}^{n_s}\left|y_{s}-\widehat{y}_{s}\right|$$
is the prediction error from the $r^{th}$ replication based on the out-of-sample observations $\{\boldsymbol{x}_{s},y_{s}\}_{s=1}^{n_s}$ that vary across the replications and a given averaging/selection method that uses $\widehat{y}_{s}$ as the predictive value.  We set $n_s=n$.
\section{Simulation}\label{sec:simu}
The purpose of this section is to evaluate, via a simulation study, the finite sample performance of our proposed methods. Note that the out-of-sample testing observations $\{\boldsymbol{x}_{s},y_{s}\}_{s=1}^{n_s}$ are no contamination in the simulation study.
For comparison, we consider some similar simulation setups as that in \cite{Agostinelli2002Robust}.
%(To assess the finite sample properties of the estimator)

\textbf{Setting A}
$$y_{i}=\nu\cdot\boldsymbol{x}_{i}^{T}\Theta+\varepsilon_{i},~~~~~~~~~~~~i=1,2,\cdots,n,$$
where $\Theta=(1, 0.1, 0, 0, 0.5, 0)^{T}$, $x_{i1}=1$ and $x_{ij}$, $j=2,\cdots,6,$ are independent and identically distributed uniform random samples in the interval $(-5,5)$, $\nu$ is varied so that $R^{2}=\frac{\mathrm{Var}(y_{i})-\mathrm{Var}(\varepsilon_{i})}{\mathrm{Var}(y_{i})}= 0.1,0.3,\cdots,0.9$.  $y_{i}$ is generated according to the model.
For the error term $\varepsilon_{i}$,
we consider normal distribution and other two different distributions to represent various deviations from normality, as described next.\\
\textbf{Case} $\boldsymbol{1}$ The error distribution follows the normal distribution
$\mathcal{N}(0,1)$, where no contamination is present.\\
\textbf{Case} $\boldsymbol{2}$ $93\%$ and $85\%$ of the samples are from a standard normal and the remaining  $7\%$ and $15\%$ from a normal with mean $0$ and standard deviation $25$.\\
\textbf{Case} $\boldsymbol{3}$ $93\%$ and $85\%$ of the samples are from a standard normal and the remaining $7\%$ and $15\%$ from a normal with mean $30$ and standard deviation $1$.

We consider candidate models with the combination of all variables and each model contains an intercept term.
Replicate $R=500$ times. Tables \ref{tab:setA1}-\ref{tab:setA3} report the $\mathrm{APEs}$ of the various estimators based on our proposed methods and the methods provided in Section \ref{sec:methods} for Case 1-3 in Setting A. The results are expressed in terms of $R^{2}$ and sample size $n=50$ and $100$.

From Tables \ref{tab:setA1}-\ref{tab:setA3}, we can see that the commonly used MMA method is superior to other methods in terms of minimizing $\mathrm{APEs}$ but is the worst when the data is contaminated. So it is very meaningful to develop robust model averaging methods.

Let $\text{A}>\text{B}$ indicate that method A performs better than B.
From Table \ref{tab:setA1}, for Case 1, of the sample size considered here, we have
$\text{MMA}>\text{MA}_{H}>\text{MA}_{H}^{c}>\text{WC}_{p}>\text{MS}_{H}>\{\text{MC}_{p},\text{HC}_{p}\}$.
While for the model averaging based on absolute loss function, the result is
$\text{MMA}>\text{WC}_{p}>\text{MA}_{A}^{c}>\{\text{MC}_{p},\text{HC}_{p}\}>\text{MS}_{A}>\text{MA}_{A}$.
When the data is not contaminated, as shown in Table \ref{tab:setA1}, we find the performance of MA$_{H}$ is slightly worse than MMA but clearly denominates the other robust model selection methods, including WC$_{p}$, MC$_{p}$ and HC$_{p}$.

When there are outlier in data,
for example the result of Case 2 in Table \ref{tab:setA2}, we can see that
$\text{WC}_{p}>\text{MA}_{H}>\text{MA}_{H}^{c}>\text{MS}_{H}>\{\text{MC}_{p},\text{HC}_{p}\}>\text{MMA}$
and
$\text{WC}_{p}>\text{MA}_{A}^{c}>\text{MS}_{A}>\{\text{MC}_{p},\text{HC}_{p}\}>\text{MA}_{A}>\text{MMA}$.

While for Case 3, when the number of outliers is $7\%$ of the sample, the performances of the above methods are the same as those in Case 2. However, when $n=50$ and the number of outliers is $15\%$ of the sample, see Table \ref{tab:setA3}, the results are different, that are
$\text{MA}_{H}>\text{MA}_{H}^{c}>\{\text{MC}_{p},\text{HC}_{p}\}>\text{MS}_{H}>\text{WC}_{p}>\text{MMA}$
and
$\text{MA}_{A}^{c}>\text{MS}_{A}>\{\text{MC}_{p},\text{HC}_{p}\}>\text{MA}_{A}>\text{WC}_{p}>\text{MMA}$.
Though WC$_{p}$ is superior to other methods for Case 2 but is inferior to others except MMA method based on the squared-error loss function for Case 3. Therefore, the performance of WC$_{p}$ is unstable.
In contrast, MA$_{H}$ and MA$_{H}^{c}$ are often in the top two. Using the absolute loss function, MA$_{A}^{c}$ performs better than MA$_{A}$ and it is also a method worth promoting.
Comparison of model averaging methods based on two different loss functions, we find the model averaging method based on Huber's loss function is more favored in most cases.
To sum up, the final conclusion is that method MA$_{H}$ is the most favored one in Setting A.
\begin{table}[b]
  \centering
  \caption{APEs of estimators for Case 1 of Setting A}
  \scalebox{0.9}{
    \begin{tabular}{crrrrrrrrrrr}
    \toprule
    n     & \multicolumn{1}{l}{$R^{2}$} & \multicolumn{1}{l}{MA$_{A}$} & \multicolumn{1}{l}{MA$_{A}^{c}$} & \multicolumn{1}{l}{MS$_{A}$} & \multicolumn{1}{l}{MA$_{H}$} & \multicolumn{1}{l}{MA$_{H}^{c}$} & \multicolumn{1}{l}{MS$_{H}$} & \multicolumn{1}{l}{WC$_{p}$} & \multicolumn{1}{l}{MC$_{p}$} & \multicolumn{1}{l}{HC$_{p}$} & \multicolumn{1}{l}{MMA} \\
    \midrule
    \multirow{5}[2]{*}{50} & 0.1   & 0.8723  & 0.8407  & 0.8571  & 0.8277  & 0.8284  & 0.8398  & 0.8376  & 0.8498  & 0.8480  & 0.8264  \\
          & 0.3   & 0.8803  & 0.8443  & 0.8543  & 0.8331  & 0.8341  & 0.8432  & 0.8409  & 0.8573  & 0.8557  & 0.8322  \\
          & 0.5   & 0.8830  & 0.8520  & 0.8632  & 0.8412  & 0.8417  & 0.8507  & 0.8501  & 0.8603  & 0.8613  & 0.8405  \\
          & 0.7   & 0.8742  & 0.8524  & 0.8643  & 0.8385  & 0.8382  & 0.8480  & 0.8442  & 0.8635  & 0.8624  & 0.8355  \\
          & 0.9   & 0.8817  & 0.8567  & 0.8634  & 0.8399  & 0.8393  & 0.8454  & 0.8433  & 0.8969  & 0.9007  & 0.8375  \\
    \midrule
    \multirow{5}[2]{*}{100} & 0.1   & 0.8339  & 0.8179  & 0.8243  & 0.8110  & 0.8116  & 0.8159  & 0.8152  & 0.8242  & 0.8239  & 0.8107  \\
          & 0.3   & 0.8349  & 0.8195  & 0.8245  & 0.8137  & 0.8142  & 0.8198  & 0.8190  & 0.8231  & 0.8221  & 0.8131  \\
          & 0.5   & 0.8304  & 0.8190  & 0.8243  & 0.8122  & 0.8123  & 0.8166  & 0.8155  & 0.8216  & 0.8214  & 0.8114  \\
          & 0.7   & 0.8386  & 0.8278  & 0.8330  & 0.8196  & 0.8198  & 0.8238  & 0.8224  & 0.8361  & 0.8364  & 0.8189  \\
          & 0.9   & 0.8394  & 0.8244  & 0.8274  & 0.8164  & 0.8168  & 0.8201  & 0.8181  & 0.8433  & 0.8482  & 0.8153  \\
    \bottomrule
    \end{tabular}}%
  \label{tab:setA1}%
\end{table}%
\begin{table}[b]
  \centering
  \caption{APEs of estimators for Case 2 of Setting A}
  \scalebox{0.85}{
    \begin{tabular}{ccrrrrrrrrrrr}
    \toprule
          & n     & \multicolumn{1}{l}{$R^{2}$} & \multicolumn{1}{l}{MA$_{A}$} & \multicolumn{1}{l}{MA$_{A}^{c}$} & \multicolumn{1}{l}{MS$_{A}$} & \multicolumn{1}{l}{MA$_{H}$} & \multicolumn{1}{l}{MA$_{H}^{c}$} & \multicolumn{1}{l}{MS$_{H}$} & \multicolumn{1}{l}{WC$_{p}$} & \multicolumn{1}{l}{MC$_{p}$} & \multicolumn{1}{l}{HC$_{p}$} & \multicolumn{1}{l}{MMA} \\
    \midrule
    \multirow{10}[4]{*}{7\%} & \multirow{5}[2]{*}{50} & 0.1   & 0.8799 & 0.8392 & 0.8563 & 0.8300  & 0.8326 & 0.8492 & 0.8333 & 0.8487 & 0.8491 & 1.5751 \\
          &       & 0.3   & 0.8944 & 0.8548 & 0.8693 & 0.8463 & 0.8493 & 0.8623 & 0.8451 & 0.8731 & 0.8720 & 1.6073 \\
          &       & 0.5   & 0.8936 & 0.8540 & 0.8713 & 0.8428 & 0.8452 & 0.8590 & 0.8422 & 0.8702 & 0.8685 & 1.6214 \\
          &       & 0.7   & 0.8871 & 0.8540 & 0.8684 & 0.8427 & 0.8448 & 0.8571 & 0.8462 & 0.8704 & 0.8695 & 1.6712 \\
          &       & 0.9   & 0.8934 & 0.8639 & 0.8746 & 0.8501 & 0.8515 & 0.8608 & 0.8486 & 0.9259 & 0.9288 & 1.7275 \\
\cmidrule{2-13}          & \multirow{5}[2]{*}{100} & 0.1   & 0.8413 & 0.8233 & 0.8279 & 0.8193 & 0.8206 & 0.8264 & 0.8199 & 0.8315 & 0.8317 & 1.2137 \\
          &       & 0.3   & 0.8402 & 0.8216 & 0.8278 & 0.8170 & 0.8182 & 0.8234 & 0.8193 & 0.8279 & 0.8275 & 1.2357 \\
          &       & 0.5   & 0.8394 & 0.8250 & 0.8321 & 0.8186 & 0.8196 & 0.8253 & 0.8202 & 0.8306 & 0.8297 & 1.2307 \\
          &       & 0.7   & 0.8415 & 0.8268 & 0.8320 & 0.8201 & 0.8207 & 0.8251 & 0.8195 & 0.8374 & 0.8368 & 1.2938 \\
          &       & 0.9   & 0.8418 & 0.8278 & 0.8312 & 0.8213 & 0.8222 & 0.8247 & 0.8208 & 0.8565 & 0.8583 & 1.2873 \\
    \midrule
    \multirow{10}[4]{*}{15\%} & \multirow{5}[2]{*}{50} & 0.1   & 0.9176 & 0.8618 & 0.882 & 0.8557 & 0.8622 & 0.8863 & 0.8568 & 0.8766 & 0.8769 & 2.0855 \\
          &       & 0.3   & 0.9120 & 0.8656 & 0.8808 & 0.8585 & 0.8637 & 0.8814 & 0.8518 & 0.8894 & 0.8876 & 2.1493 \\
          &       & 0.5   & 0.9178 & 0.8664 & 0.8826 & 0.8617 & 0.8672 & 0.8850 & 0.8527 & 0.8893 & 0.8877 & 2.1338 \\
          &       & 0.7   & 0.9154 & 0.8764 & 0.8932 & 0.8705 & 0.8755 & 0.8944 & 0.8669 & 0.9000 & 0.9049 & 2.1513 \\
          &       & 0.9   & 0.9117 & 0.8728 & 0.8841 & 0.8665 & 0.8720 & 0.8825 & 0.8464 & 0.9559 & 0.9511 & 2.3456 \\
\cmidrule{2-13}          & \multirow{5}[2]{*}{100} & 0.1   & 0.8576 & 0.8345 & 0.8439 & 0.8295 & 0.8317 & 0.8407 & 0.8294 & 0.8430 & 0.8431 & 1.5094 \\
          &       & 0.3   & 0.8545 & 0.8322 & 0.8399 & 0.8283 & 0.8306 & 0.8383 & 0.8276 & 0.8402 & 0.8407 & 1.5196 \\
          &       & 0.5   & 0.8492 & 0.8310 & 0.8391 & 0.8260 & 0.8278 & 0.8355 & 0.8211 & 0.8395 & 0.8394 & 1.5519 \\
          &       & 0.7   & 0.8565 & 0.8385 & 0.8451 & 0.8332 & 0.8348 & 0.8416 & 0.8267 & 0.8550 & 0.8551 & 1.5776 \\
          &       & 0.9   & 0.8496 & 0.8308 & 0.8330 & 0.8265 & 0.8282 & 0.8332 & 0.8211 & 0.8795 & 0.8812 & 1.5931 \\
    \bottomrule
    \end{tabular}}%
  \label{tab:setA2}%
\end{table}%
\begin{table}[b]
  \centering
  \caption{APEs of estimators for Case 3 of Setting A}
  \scalebox{0.85}{
    \begin{tabular}{ccrrrrrrrrrrr}
    \toprule
          & n     & \multicolumn{1}{l}{$R^{2}$} & \multicolumn{1}{l}{MA$_{A}$} & \multicolumn{1}{l}{MA$_{A}^{c}$} & \multicolumn{1}{l}{MS$_{A}$} & \multicolumn{1}{l}{MA$_{H}$} & \multicolumn{1}{l}{MA$_{H}^{c}$} & \multicolumn{1}{l}{MS$_{H}$} & \multicolumn{1}{l}{WC$_{p}$} & \multicolumn{1}{l}{MC$_{p}$} & \multicolumn{1}{l}{HC$_{p}$} & \multicolumn{1}{l}{MMA} \\
    \midrule
    \multirow{10}[4]{*}{7\%} & \multirow{5}[2]{*}{50} & 0.1   & 0.8883  & 0.8437  & 0.8590  & 0.8407  & 0.8440  & 0.8602  & 0.8357  & 0.8597  & 0.8589  & 2.6323  \\
          &       & 0.3   & 0.9000  & 0.8590  & 0.8722  & 0.8538  & 0.8572  & 0.8697  & 0.8466  & 0.8804  & 0.8793  & 2.6573  \\
          &       & 0.5   & 0.8982  & 0.8582  & 0.8724  & 0.8523  & 0.8557  & 0.8698  & 0.8484  & 0.8730  & 0.8757  & 2.6544  \\
          &       & 0.7   & 0.8983  & 0.8680  & 0.8801  & 0.8584  & 0.8611  & 0.8724  & 0.8467  & 0.8874  & 0.8891  & 2.7271  \\
          &       & 0.9   & 0.8951  & 0.8691  & 0.8784  & 0.8591  & 0.8613  & 0.8678  & 0.8424  & 0.9413  & 0.9404  & 2.7558  \\
\cmidrule{2-13}          & \multirow{5}[2]{*}{100} & 0.1   & 0.8473  & 0.8277  & 0.8336  & 0.8264  & 0.8276  & 0.8328  & 0.8194  & 0.8388  & 0.8385  & 2.2219  \\
          &       & 0.3   & 0.8477  & 0.8270  & 0.8332  & 0.8259  & 0.8272  & 0.8324  & 0.8196  & 0.8363  & 0.8371  & 2.2394  \\
          &       & 0.5   & 0.8446  & 0.8292  & 0.8365  & 0.8269  & 0.8277  & 0.8334  & 0.8203  & 0.8359  & 0.8355  & 2.2559  \\
          &       & 0.7   & 0.8476  & 0.8373  & 0.8416  & 0.8343  & 0.8346  & 0.8384  & 0.8242  & 0.8529  & 0.8521  & 2.2782  \\
          &       & 0.9   & 0.8462  & 0.8313  & 0.8337  & 0.8268  & 0.8276  & 0.8308  & 0.8205  & 0.8633  & 0.8647  & 2.2535  \\
    \midrule
    \multirow{10}[4]{*}{15\%} & \multirow{5}[2]{*}{50} & 0.1   & 0.9513  & 0.8893  & 0.9107  & 0.9240  & 0.9325  & 0.9591  & 1.0566  & 0.9276  & 0.9263  & 4.8421  \\
          &       & 0.3   & 0.9615  & 0.9059  & 0.9140  & 0.9425  & 0.9523  & 0.9693  & 1.0818  & 0.9514  & 0.9499  & 4.8670  \\
          &       & 0.5   & 0.9495  & 0.8952  & 0.9047  & 0.9341  & 0.9462  & 0.9622  & 1.1985  & 0.9403  & 0.9433  & 4.9460  \\
          &       & 0.7   & 0.9465  & 0.9047  & 0.9163  & 0.9439  & 0.9534  & 0.9736  & 1.1561  & 0.9617  & 0.9627  & 4.9474  \\
          &       & 0.9   & 0.9430  & 0.9186  & 0.9208  & 0.9592  & 0.9679  & 0.9729  & 1.2011  & 1.0383  & 1.0446  & 4.9718  \\
\cmidrule{2-13}          & \multirow{5}[2]{*}{100} & 0.1   & 0.8774  & 0.8500  & 0.8564  & 0.8788  & 0.8820  & 0.8909  & 0.8191  & 0.8783  & 0.8786  & 4.5344  \\
          &       & 0.3   & 0.8713  & 0.8478  & 0.8536  & 0.8757  & 0.8793  & 0.8887  & 0.8204  & 0.8783  & 0.8785  & 4.5399  \\
          &       & 0.5   & 0.8719  & 0.8495  & 0.8568  & 0.8790  & 0.8818  & 0.8913  & 0.8204  & 0.8828  & 0.8817  & 4.5183  \\
          &       & 0.7   & 0.8751  & 0.8605  & 0.8675  & 0.8881  & 0.8903  & 0.8987  & 0.8246  & 0.8963  & 0.8984  & 4.5456  \\
          &       & 0.9   & 0.8785  & 0.8634  & 0.8620  & 0.8881  & 0.8908  & 0.8956  & 0.8201  & 0.9366  & 0.9360  & 4.5511  \\
    \bottomrule
    \end{tabular}}%
  \label{tab:setA3}%
\end{table}%

In the following, we take the Example $1$ from \cite{Ronchetti1994A}, specific as follows.

\textbf{Setting B}
$$y_{i}=x_{i1}+x_{i2}+\varepsilon_{i},~~~~~~~~~~~~i=1,2,\cdots,20,$$
where $x_{i1}$ and $x_{i2}$ are independent and generated from uniform distribution on $(-1,1)$；
$\varepsilon_{i}$ are independent normally distributed errors with expectation $0$ and standard deviation $\sigma=0.3$, $1$ and $2$.
One more uniform random variable $x_{i3}$ in the interval $(-1,1)$ is also considered in the
study as possible explanatory variable. Therefore, the candidate models are constructed as Setting A then we obtain
$7$ alternative models.
We implement both classical and robust procedures on this data (case I) and on the same data with the point $y_{20}$
changed to $10$ (case II). Replicate $R=500$ times.

From the results of Setting B in Table \ref{tab:setB}, we can find more information that is not presented in Setting A. Note that when $\sigma=0.3$, the robust model selection methods WC$_{p}$, MC$_{p}$ and HC$_{p}$ are
superior to our robust model averaging methods in most cases, but as $\sigma$ increases, MA$_{H}$, MA$_{H}^{c}$ and MA$_{A}^{c}$ perform much better than the model selection methods, such as when $\sigma=1.5$ in Table \ref{tab:setB}
In conclusion, MA$_{H}$ is found that a high level of model noise stability can be achieved.

All simulation results imply that MA$_{H}$ significantly
out-performs other model averaging and model selection methods in achieving the lowest APEs for outlier pollution data, a promising and meaningful result.

\begin{table}[b]
  \centering
  \caption{APEs of estimators for Setting B}
  \scalebox{0.9}{
    \begin{tabular}{crrrrrrrrrrr}
    \toprule
          & \multicolumn{1}{l}{$\sigma$} & \multicolumn{1}{l}{MA$_{A}$} & \multicolumn{1}{l}{MA$_{A}^{c}$} & \multicolumn{1}{l}{MS$_{A}$} & \multicolumn{1}{l}{MA$_{H}$} & \multicolumn{1}{l}{MA$_{H}^{c}$} & \multicolumn{1}{l}{MS$_{H}$} & \multicolumn{1}{l}{WC$_{p}$} & \multicolumn{1}{l}{MC$_{p}$} & \multicolumn{1}{l}{HC$_{p}$} & \multicolumn{1}{l}{MMA} \\
    \midrule
    \multirow{3}[2]{*}{Case I} & 0.3   & 0.2788  & 0.2722  & 0.2742  & 0.2645  & 0.2642  & 0.2658  & 0.2656  & 0.2716  & 0.2726  & 0.2637  \\
          & 1     & 0.9260  & 0.9162  & 0.9390  & 0.8869  & 0.8853  & 0.8956  & 0.8989  & 0.9315  & 0.9279  & 0.8840  \\
          & 1.5   & 1.3850  & 1.3522  & 1.3921  & 1.3165  & 1.3169  & 1.3471  & 1.3450  & 1.3610  & 1.3604  & 1.3099  \\
    \midrule
    \multirow{3}[2]{*}{Case II} & 0.3   & 0.2851  & 0.2787  & 0.2793  & 0.2820  & 0.2905  & 0.2716  & 0.2658  & 0.2749  & 0.2755  & 0.7934  \\
          & 1     & 0.9624  & 0.9481  & 0.9757  & 0.9190  & 0.9201  & 0.9360  & 0.9238  & 0.9645  & 0.9627  & 1.1366  \\
          & 1.5   & 1.4254  & 1.3799  & 1.4239  & 1.3458  & 1.3490  & 1.3857  & 1.3687  & 1.3972  & 1.3947  & 1.4878  \\
    \bottomrule
    \end{tabular}}%
  \label{tab:setB}%
\end{table}%
\section{Data examples}\label{sec:data}
In this section, we present three data examples to evaluate the finite sample performance of our proposed robust model average estimators.
The candidate models setting is same as that in the simulation study.
To illustrate the new methods,
suppose we know that the samples $\{y_{o_{i}}\}_{i=1}^{n_{1}}$ are outliers. The remaining samples are $\{y_{t_{i}}\}_{i=1}^{n_{2}}$. We adopt a delete-one prediction method.
For example, we delete $\{\boldsymbol{x}_{t_{1}},y_{t_{1}}\}$,
while we use samples $\{y_{o_{i}}\}_{i=1}^{n_{1}}$ and $\{y_{t_{i}}\}_{i=2}^{n_{2}}$ as training samples and to predict $y_{t_{1}}$ by all robust methods we considered. $\widehat{y}_{t}$ is denoted as the predicted value of $y_{t}$.
We repeat the above steps for $\{\boldsymbol{x}_{t_{i}},y_{t_{i}}\}_{i=2}^{n_{2}}$. Finally, We get the predicted values $\{\widehat{y}_{t_{i}}\}_{i=1}^{n_{2}}$.
Then we evaluate all robust methods with respect to the predict error $\mathrm{APE}=\sum_{i=1}^{n_{2}}|y_{t_{i}}-\widehat{y}_{t_{i}}|/n_{2}$.
\subsection{Outlier polluted data}
\subsubsection{Artificial data}
The first data example is the artificial data set generated by \cite{Hawkins1984Location} and can be obtained from R software. The data set consists of $75$ observations in four dimensions (one response and three explanatory variables).
It provides a good example of the masking effect. The first $14$ observations are outliers, created in two groups: $1$-$10$ and $11$-$14$.
%Only observations $12$, $13$ and $14$ appear as outliers when using classical methods.

%%%%%%%%%%%%%%%%%%%%%%%%%%%%%%%%%%%%%%%%%%%%%%%%%%%%%%%%%%%%%%%%%%%%%%%%%%%%%
%
\begin{table}[htbp]
  \centering
  \caption{APEs of estimators for the Artificial data example}
  \vskip2mm
    \begin{tabular}{rrrrrrrrrr}
    \toprule
    \multicolumn{1}{l}{MA$_{A}$} & \multicolumn{1}{l}{MA$_{A}^{c}$} & \multicolumn{1}{l}{MS$_{A}$} & \multicolumn{1}{l}{MA$_{H}$} & \multicolumn{1}{l}{MA$_{H}^{c}$} & \multicolumn{1}{l}{MS$_{H}$} & \multicolumn{1}{l}{WC$_{p}$} & \multicolumn{1}{l}{MC$_{p}$} & \multicolumn{1}{l}{HC$_{p}$} & \multicolumn{1}{l}{MMA} \\
   \midrule
    0.5891  & 0.6122  & 0.6278  & 0.6102  & 0.6087  & 0.6144  & 0.5135  & 0.6945  & 0.7643  & 0.6674  \\
    \bottomrule
    \end{tabular}%
  \label{tab:Shkb}%
\end{table}%

\subsubsection{Data of conversion of Ammonia to Nitric Acid  }
\cite{Andrews1974A} discusses a set of data that are most interesting
from the point of view of outlier detection. The data with $21$ observations and $3$ independent variables, which is
reproduced in Table \ref{tab:Acid}, relates to the conversion of ammonia to nitric acid. Stack Loss is the response variable and others are predictors. \cite{Hawkins1980} asserted that a conventional probability plot of residuals suggests that observation $21$ may be an outlier. Thus, we can use this data to evaluate our methods.
\begin{table}[b]
  \centering
  \caption{Data from Operation of A Plant for the Oxidation of Ammonia to Nitric Acid}
  \vskip2mm
  \scalebox{0.9}{
    \begin{tabular}{ccccc}
    \toprule
    \multirow{2}[2]{*}{Observation Number} & \multirow{2}[2]{*}{Stack Loss} & \multirow{2}[2]{*}{Air Flow} & Cooling Water  & \multirow{2}[2]{*}{Acid Concentration} \\
          &       &       & Inlet Temperature &  \\
    \midrule
    1     & 42    & 80    & 27    & 89 \\
    2     & 37    & 80    & 27    & 88 \\
    3     & 37    & 75    & 25    & 90 \\
    4     & 28    & 62    & 24    & 87 \\
    5     & 18    & 62    & 22    & 87 \\
    6     & 18    & 62    & 23    & 87 \\
    7     & 19    & 62    & 24    & 93 \\
    8     & 20    & 62    & 24    & 93 \\
    9     & 15    & 58    & 23    & 87 \\
    10    & 14    & 58    & 18    & 80 \\
    11    & 14    & 58    & 18    & 89 \\
    12    & 13    & 58    & 17    & 88 \\
    13    & 11    & 58    & 18    & 82 \\
    14    & 12    & 58    & 19    & 93 \\
    15    & 8     & 50    & 18    & 89 \\
    16    & 7     & 50    & 18    & 86 \\
    17    & 8     & 50    & 19    & 72 \\
    18    & 8     & 50    & 19    & 79 \\
    19    & 9     & 50    & 20    & 80 \\
    20    & 15    & 56    & 20    & 82 \\
    21    & 15    & 70    & 20    & 91 \\
    \bottomrule
    \end{tabular}}%
  \label{tab:Acid}%
\end{table}%
\begin{table}[htbp]
  \centering
  \caption{APEs of estimators for the data of conversion of Ammonia to Nitric Acid}
  \vskip2mm
    \begin{tabular}{rrrrrrrrrr}
    \toprule
    \multicolumn{1}{l}{MA$_{A}$} & \multicolumn{1}{l}{MA$_{A}^{c}$} & \multicolumn{1}{l}{MS$_{A}$} & \multicolumn{1}{l}{MA$_{H}$} & \multicolumn{1}{l}{MA$_{H}^{c}$} & \multicolumn{1}{l}{MS$_{H}$} & \multicolumn{1}{l}{WC$_{p}$} & \multicolumn{1}{l}{MC$_{p}$} & \multicolumn{1}{l}{HC$_{p}$} & \multicolumn{1}{l}{MMA} \\
    \midrule
    1.6904  & 2.0561  & 1.9571  & 2.1576  & 2.1162  & 2.7303  & 2.8259  & 2.2715  & 2.4940  & 2.4780  \\
    \bottomrule
    \end{tabular}%
  \label{tab:SAcid}%
\end{table}%

These two examples are outlier polluted data. The $\mathrm{APEs}$ results are presented in Tables \ref{tab:Shkb} and \ref{tab:SAcid}.
For the Artificial data, the prediction result is $\text{WC}_{p}>\text{MA}_{H}^{c}>\text{MA}_{H}>\text{MS}_{H}>\text{MMA}>\{\text{MC}_{p},\text{HC}_{p}\}$,
while for the model averaging based on absolute loss function, the result is
$\text{WC}_{p}>\text{MA}_{A}>\text{MA}_{A}^{c}>\text{MS}_{A}>\text{MMA}>\{\text{MC}_{p},\text{HC}_{p}\}$.
We find that the performances of our methods are still very good, in the forefront.
Though WC$_{p}$ is superior to other methods for this example but is inferior to others for the second example. In Table \ref{tab:SAcid}, the results are
$\text{MA}_{H}^{c}>\text{MA}_{H}>\{\text{MC}_{p},\text{HC}_{p},\text{MMA}\}>\text{MS}_{H}>\text{WC}_{p}$
and
$\text{MA}_{A}>\text{MS}_{A}>\text{MA}_{A}^{c}>\{\text{MC}_{p},\text{HC}_{p},\text{MMA}\}>\text{WC}_{p}$.
WC$_{p}$ is the least favored method, indicating the unstable of this method again.
The model averaging method based on absolute loss function is more favored in most cases, which is different from the result in the simulation study. But the main trend is still that
the proposed methods, such as MA$_{A}$ and MA$_{H}^{c}$, are often in the top two for these two examples.
\subsection{The Hald Cement Data}
We investigate the Hald cement data provided in \cite{Ronchetti1994A}, which are shown in Table \ref{tab:Cement}. It is not an outlier polluted data but some variables are highly correlated.
The response variable is the heat evolved $y$ in a cement mix, and the four explanatory variables are ingredients in
the mix.
\cite{Ronchetti1994A} found that the residuals show no evidence of any problems when a linear model is fitted,
but an important feature of these data is that the variables $x_{1}$ and $x_{3}$ are highly correlated.

The $\mathrm{APEs}$ results of these methods are provided in Table \ref{tab:SCement},
implying that MA$_{A}$ and MA$_{H}$ always yields the best two results.
We see that
$\text{MA}_{H}>\text{MA}_{H}^{c}>\{\text{MC}_{p}, \text{HC}_{p}, \text{MMA}\}>\text{WC}_{p}>\text{MS}_{H}$
and
$\text{MA}_{A}>\text{MS}_{A}>\text{MA}_{A}^{c}>\{\text{MC}_{p}, \text{HC}_{p}, \text{MMA}\}>\text{WC}_{p}$.
All of our methods are still perform better than MMA.
The results of this example show that our method is also robust to covariate highly correlated data.
Mallows-type model averaging methods based on absolute and Huber's loss functions are worthy of advocating.

\begin{table}[b]
  \centering
  \caption{The Hald Cement Data}
  \vskip2mm
  \scalebox{0.9}{
    \begin{tabular}{cccccc}
    \toprule
    ~~~$i$~~~   & ~~~$y_{i}$~~~ & ~~~$x_{i1}$~~~ & ~~~$x_{i2}$~~~ & ~~~$x_{i3}$~~~ & ~~~$x_{i4}$~~~ \\
    \midrule
    1     & 78.5  & 7     & 26    & 6     & 60 \\
    2     & 74.3  & 1     & 29    & 15    & 52 \\
    3     & 104.3 & 11    & 56    & 8     & 20 \\
    4     & 87.6  & 11    & 31    & 8     & 47 \\
    5     & 95.9  & 7     & 52    & 6     & 33 \\
    6     & 109.2 & 11    & 55    & 9     & 22 \\
    7     & 102.7 & 3     & 71    & 17    & 6 \\
    8     & 72.5  & 1     & 31    & 22    & 44 \\
    9     & 93.1  & 2     & 54    & 18    & 22 \\
    10    & 115.9 & 21    & 47    & 4     & 26 \\
    11    & 83.8  & 1     & 40    & 23    & 34 \\
    12    & 113.3 & 11    & 66    & 9     & 12 \\
    13    & 109.4 & 10    & 68    & 8     & 12 \\
    \bottomrule
    \end{tabular}}%
  \label{tab:Cement}%
\end{table}%
\begin{table}[htbp]
  \centering
  \caption{APEs of estimators for the Hald cement data}
  \vskip2mm
    \begin{tabular}{rrrrrrrrrr}
    \toprule
     \multicolumn{1}{l}{MA$_{A}$} & \multicolumn{1}{l}{MA$_{A}^{c}$} & \multicolumn{1}{l}{MS$_{A}$} & \multicolumn{1}{l}{MA$_{H}$} & \multicolumn{1}{l}{MA$_{H}^{c}$} & \multicolumn{1}{l}{MS$_{H}$} & \multicolumn{1}{l}{WC$_{p}$} & \multicolumn{1}{l}{MC$_{p}$} & \multicolumn{1}{l}{HC$_{p}$} & \multicolumn{1}{l}{MMA} \\
    \midrule
    2.0640  & 2.4286  & 2.3180  & 2.5372  & 2.5116  & 3.0952  & 2.9915  & 2.7933  & 2.8600  & 2.8550  \\
    \bottomrule
    \end{tabular}%
  \label{tab:SCement}%
\end{table}%
\subsection{Aerobic Fitness Prediction}
Like \cite{Agostinelli2002Robuststep}, we consider the dataset from the SAS/STAT User's Guide (1990, p. 1443) with 31 observations.
There are one dependent variable and and six explanatory variables, including oxygen intake rate (ml per kg of body weight per minute), time to run 1.5 miles (minutes), age (year), weight (kg), heart rate while running (at the same time as oxygen rate measured), maximum heart rate recorded while running and  heart rate while resting. \cite{Agostinelli2002Robuststep} found observation 10 is a moderate leverage point.

The $\mathrm{APEs}$ results of these methods are provided in Table \ref{tab:Fitness},
implying that
$\text{MA}_{H}^{c}>\text{MS}_{H}>\text{MA}_{H}>\text{MMA}>\text{WC}_{p}>\{\text{MC}_{p},\text{HC}_{p}\}$,
while for the model averaging based on absolute loss function, the result is
$\text{MS}_{A}>\text{MA}_{A}>\text{MA}_{A}^{c}>\text{MMA}>\text{WC}_{p}>\{\text{MC}_{p},\text{HC}_{p}\}$.
The results of this example show that our method is also robust to the data with leverage point.
\begin{table}[htbp]
  \centering
  \caption{APEs of estimators for the Aerobic Fitness Prediction}
  \vskip2mm
    \begin{tabular}{rrrrrrrrrr}
    \toprule
     \multicolumn{1}{l}{MA$_{A}$} & \multicolumn{1}{l}{MA$_{A}^{c}$} & \multicolumn{1}{l}{MS$_{A}$} & \multicolumn{1}{l}{MA$_{H}$} & \multicolumn{1}{l}{MA$_{H}^{c}$} & \multicolumn{1}{l}{MS$_{H}$} & \multicolumn{1}{l}{WC$_{p}$} & \multicolumn{1}{l}{MC$_{p}$} & \multicolumn{1}{l}{HC$_{p}$} & \multicolumn{1}{l}{MMA} \\
     \midrule
    1.5875  & 1.5920  & 1.5376  & 1.6664  & 1.6355  & 1.6628  & 1.9888  & 30.1219  & 58.9784  & 1.8513  \\
    \bottomrule
    \end{tabular}%
  \label{tab:Fitness}%
\end{table}%
\section{Concluding remarks}\label{sec:conc}
Model averaging is an alternative to model selection for dealing with model uncertainty, which is widely used and very valuable.
However, most of the existing model averaging methods are proposed based on the squared loss function, which could be very sensitive to the presence of outliers in the data.
Widely used model averaging criteria, such as the Mallows criterion proposed in \cite{hansen2007least, hansen2008joe, wan2010least}, have favourable properties if their underlying conditions are true, but misleading results if they are not true.
In this paper, we have proposed an outlier-robust model averaging approach by Mallows-type criterion. The key idea is to develop weight choice criteria by minimising an estimate of the expected prediction error for the function being convex with a unique minimum, and twice differentiable in expectation, rather than the
expected squared error. The robust loss functions, such as least absolute deviation and Huber's function, reduce the effects of large residuals and poor samples. Both simulation study and actual data analysis favor our robust model averaging approach.

In the future, we can study the robust model averaging methods based on other criteria, such as cross-validation procedure.
In this context, we only have deduced the criteria, and in the future we shall prove some excellent properties of these robust average weight estimators, such as the asymptotic unbiasedness of them.

%%%%%%%%%%%%%%%%%%%%%%%%%%%%%%%%%%%%%%%%%%%%%%%%%%%%%%%%%%%%%%%%%%%%%%%%%%%%%%%%%%%%%%
%%%%%%%%%%%%%%%%%%%%%%%%%%%%%%%%%%%%%%%%%%%%%%%%%%%%%%%%%%%%%%%%%%%%%%%%%%%%%%%%%%%%%%
\appendix
\renewcommand\thesection{\appendixname~\Alph{section}}
\renewcommand\theequation{\Alph{section}.\arabic{equation}}
\renewcommand\thelem{\Alph{section}.\arabic{lem}}
\section{The derivation of \eqref{C:fixed}}\label{sec:A}
%\subsection{Fixed Design Matrix }
By the definition of $\Theta_{(m)}^{*}$ and Taylor expansion, for $\Theta_{(m)}$ near $\Theta_{(m)}^{*}$,
we have
\be\label{eq:rhom}
&&\sum_{i=1}^{n}\rho\left(y_{i}-\boldsymbol{x}_{i(m)}^{T}\Theta_{(m)}\right)\non\\
&=&\sum_{i=1}^{n}\rho\left(y_{i}-\boldsymbol{x}_{i(m)}^{T}\Theta_{(m)}^{*}\right)+\left(\Theta_{(m)}-\Theta_{(m)}^{*}\right)^{T}\sum_{i=1}^{n}\boldsymbol{x}_{i(m)}\rho_{1}\left(y_{i}-\boldsymbol{x}_{i(m)}^{T}\Theta_{(m)}^{*}\right)\non\\
&&+\frac{1}{2}\left(\Theta_{(m)}-\Theta_{(m)}^{*}\right)^{T}
\sum_{i=1}^{n}\boldsymbol{x}_{i(m)}R_{2}\left(\mu_{i}-\boldsymbol{x}_{i(m)}^{T}\Theta_{(m)}^{*}\right)
\boldsymbol{x}_{i(m)}^{T}\left(\Theta_{(m)}-\Theta_{(m)}^{*}\right)\non\\
&&\cdot\left(1+o(1)\right)+o_{p}(1).
\ee
By the definition of $\widehat{\Theta}_{(m)}$ and \eqref{eq:rhom}, note that $\widehat{\Theta}_{(m)}$ minimises a quadratic in $\Theta_{(m)}-\Theta_{(m)}^{*}$. Let $A_{(m)}=\sum_{i=1}^{n}R_{2}\left(\mu_{i}-\boldsymbol{x}_{i(m)}^{T}\Theta_{(m)}^{*}\right)\boldsymbol{x}_{i(m)}\boldsymbol{x}_{i(m)}^{T}$ and $V_{(m)}=\sum_{i=1}^{n}\rho_{1}\left(y_{i}-\boldsymbol{x}_{i(m)}^{T}\Theta_{(m)}^{*}\right)\boldsymbol{x}_{i(m)}$.
So we can have
\be\label{eq:Theatm}
\widehat{\Theta}_{(m)}-\Theta_{(m)}^{*}&=&-A_{(m)}^{-1}
\sum_{i=1}^{n}\rho_{1}\left(y_{i}-\boldsymbol{x}_{i(m)}^{T}\Theta_{(m)}^{*}\right)\boldsymbol{x}_{i(m)}+o_{p}(1)\non\\
&=&-A_{(m)}^{-1}
V_{(m)}+o_{p}(1).
\ee

Further, by Taylor Theorem, we obtain that
\be\label{eq:rho}
&&\sum_{i=1}^{n}\rho\left(y_{i}-\sum_{m=1}^{M}w_{m}\boldsymbol{x}_{i(m)}^{T}\Theta_{(m)}\right)\non\\
&=&\sum_{i=1}^{n}\rho\left(y_{i}-\sum_{m=1}^{M}w_{m}\boldsymbol{x}_{i(m)}^{T}\Theta_{(m)}^{*}\right)\non\\
&&+\sum_{m=1}^{M}w_{m}\left(\Theta_{(m)}-\Theta_{(m)}^{*}\right)^{T}\sum_{i=1}^{n}\boldsymbol{x}_{i(m)}\rho_{1}\left(y_{i}-\sum_{m=1}^{M}w_{m}\boldsymbol{x}_{i(m)}^{T}\Theta_{(m)}^{*}\right)\non\\
&&+\frac{1}{2}\left[\sum_{m=1}^{M}w_{m}\boldsymbol{x}_{i(m)}^{T}\left(\Theta_{(m)}-\Theta_{(m)}^{*}\right)\right]^{T}
R_{2}\left(\mu_{i}-\sum_{m=1}^{M}w_{m}\boldsymbol{x}_{i(m)}^{T}\Theta_{(m)}^{*}\right)
\non\\
&&\left[\sum_{m=1}^{M}w_{m}\boldsymbol{x}_{i(m)}^{T}\left(\Theta_{(m)}-\Theta_{(m)}^{*}\right)\right]\{1+o(1)\}+o_{p}(1).
\ee

By \eqref{eq:Theatm} and \eqref{eq:rho}, we have
\be\label{eq:rhomhat}
&&\sum_{i=1}^{n}\rho\left(y_{i}-\sum_{m=1}^{M}w_{m}\boldsymbol{x}_{i(m)}^{T}\widehat{\Theta}_{(m)}\right)\non\\
&=&\sum_{i=1}^{n}\rho\left(y_{i}-\sum_{m=1}^{M}w_{m}\boldsymbol{x}_{i(m)}^{T}\Theta_{(m)}^{*}\right)\non\\
&&+\sum_{m=1}^{M}w_{m}\left(-A_{(m)}^{-1}
V_{(m)}\right)^{T}\sum_{i=1}^{n}\boldsymbol{x}_{i(m)}\rho_{1}\left(y_{i}-\sum_{m=1}^{M}w_{m}\boldsymbol{x}_{i(m)}^{T}\Theta_{(m)}^{*}\right)\non\\
&&+\frac{1}{2}\left[\sum_{m=1}^{M}w_{m}\boldsymbol{x}_{i(m)}^{T}\left(-A_{(m)}^{-1}
V_{(m)}\right)\right]^{T}
R_{2}\left(\mu_{i}-\sum_{m=1}^{M}w_{m}\boldsymbol{x}_{i(m)}^{T}\Theta_{(m)}^{*}\right)
\non\\
&&\left[\sum_{m=1}^{M}w_{m}\boldsymbol{x}_{i(m)}^{T}\left(-A_{(m)}^{-1}
V_{(m)}\right)\right]\{1+o(1)\}+o_{p}(1).
\ee
Taking expectation over $\widetilde{y}$ and the notation denoted by $\mathrm{\widetilde{E}}$, we obtain that
\be \label{f8}
&&\mathrm{\widetilde{E}}\sum_{i=1}^{n}\rho\left(\widetilde{y}_{i}-\sum_{m=1}^{M}w_{m}\boldsymbol{x}_{i(m)}^{T}\widehat{\Theta}_{(m)}\right)\non\\
&=&\sum_{i=1}^{n}\mathrm{\widetilde{E}}\rho\left(\widetilde{y}_{i}-\sum_{m=1}^{M}w_{m}\boldsymbol{x}_{i(m)}^{T}\Theta_{(m)}^{*}\right)\non\\
&&+\sum_{m=1}^{M}w_{m}\left(\widehat{\Theta}_{(m)}-\Theta_{(m)}^{*}\right)^{T}\sum_{i=1}^{n}\boldsymbol{x}_{i(m)}\mathrm{\widetilde{E}}\rho_{1}\left(\widetilde{y}_{i}-\sum_{m=1}^{M}w_{m}\boldsymbol{x}_{i(m)}^{T}\Theta_{(m)}^{*}\right)\non\\
&&+\frac{1}{2}\left[\sum_{m=1}^{M}w_{m}\boldsymbol{x}_{i(m)}^{T}\left(\widehat{\Theta}_{(m)}-\Theta_{(m)}^{*}\right)\right]^{T}
R_{2}\left(\mu_{i}-\sum_{m=1}^{M}w_{m}\boldsymbol{x}_{i(m)}^{T}\Theta_{(m)}^{*}\right)\non\\
&&\left[\sum_{m=1}^{M}w_{m}\boldsymbol{x}_{i(m)}^{T}\left(\widehat{\Theta}_{(m)}-\Theta_{(m)}^{*}\right)\right]
\{1+o(1)\}.
\ee

Taking expectation for \eqref{eq:rhomhat},
\be\label{eq:Erhomhat}
&&\mathrm{E}\sum_{i=1}^{n}\rho\left(y_{i}-\sum_{m=1}^{M}w_{m}\boldsymbol{x}_{i(m)}^{T}\widehat{\Theta}_{(m)}\right)\non\\
&=&\mathrm{E}\sum_{i=1}^{n}\rho\left(y_{i}-\sum_{m=1}^{M}w_{m}\boldsymbol{x}_{i(m)}^{T}\Theta_{(m)}^{*}\right)\non\\
&&+\mathrm{E}\left\{\sum_{m=1}^{M}w_{m}\left(-A_{(m)}^{-1}
V_{(m)}\right)^{T}\sum_{i=1}^{n}\boldsymbol{x}_{i(m)}\rho_{1}\left(y_{i}-\sum_{m=1}^{M}w_{m}\boldsymbol{x}_{i(m)}^{T}\Theta_{(m)}^{*}\right)\right\}\non\\
&&+\frac{1}{2}\mathrm{E}\left[\left\{\sum_{m=1}^{M}w_{m}\boldsymbol{x}_{i(m)}^{T}\left(-A_{(m)}^{-1}
V_{(m)}\right)\right\}^{T}
R_{2}\left(\mu_{i}-\sum_{m=1}^{M}w_{m}\boldsymbol{x}_{i(m)}^{T}\Theta_{(m)}^{*}\right)\right.\non\\
&&\left.\left\{\sum_{m=1}^{M}w_{m}\boldsymbol{x}_{i(m)}^{T}\left(-A_{(m)}^{-1}
V_{(m)}\right)\right\}\right]
\{1+o(1)\}+o_{p}(1).
\ee
Similarly, taking expectation for \eqref{f8},
\be
&&\mathrm{E}\sum_{i=1}^{n}\rho\left(\widetilde{y}_{i}-\sum_{m=1}^{M}w_{m}\boldsymbol{x}_{i(m)}^{T}\widehat{\Theta}_{(m)}\right)\non\\
&=&\mathrm{E}\sum_{i=1}^{n}\rho\left(y_{i}-\sum_{m=1}^{M}w_{m}\boldsymbol{x}_{i(m)}^{T}\widehat{\Theta}_{(m)}\right)\non\\
&&+\mathrm{E}\left\{\sum_{m=1}^{M}w_{m}\left(-A_{(m)}^{-1}
V_{(m)}\right)^{T}\right\}\sum_{i=1}^{n}\boldsymbol{x}_{i(m)}\mathrm{E}\rho_{1}\left(y_{i}-\sum_{m=1}^{M}w_{m}\boldsymbol{x}_{i(m)}^{T}\Theta_{(m)}^{*}\right)\non\\
&&+\mathrm{E}\left[\left\{\sum_{m=1}^{M}w_{m}\boldsymbol{x}_{i(m)}^{T}\left(-A_{(m)}^{-1}
V_{(m)}\right)\right\}^{T}
R_{2}\left(\mu_{i}-\sum_{m=1}^{M}w_{m}\boldsymbol{x}_{i(m)}^{T}\Theta_{(m)}^{*}\right)\right.\non\\
&&\left.\left\{\sum_{m=1}^{M}w_{m}\boldsymbol{x}_{i(m)}^{T}\left(-A_{(m)}^{-1}
V_{(m)}\right)\right\}\right]
\{1+o(1)\}.
\ee
Let $\widetilde{V}_{(m)}=\sum_{i=1}^{n}\rho_{1}\left(y_{i}-\sum_{m=1}^{M}w_{m}\boldsymbol{x}_{i(m)}^{T}\Theta_{(m)}^{*}\right)\boldsymbol{x}_{i(m)}$.
The above two equations imply
\be
&&\mathrm{E}\sum_{i=1}^{n}\rho\left(\widetilde{y}_{i}-\sum_{m=1}^{M}w_{m}\boldsymbol{x}_{i(m)}^{T}\widehat{\Theta}_{(m)}\right)\non\\
&\simeq&\sum_{i=1}^{n}\mathrm{E}\rho\left(y_{i}-\sum_{m=1}^{M}w_{m}\boldsymbol{x}_{i(m)}^{T}\widehat{\Theta}_{(m)}\right)\non\\
&&+\mathrm{E}\left\{\sum_{m=1}^{M}w_{m}\left(A_{(m)}^{-1}
V_{(m)}\right)^{T}\sum_{i=1}^{n}\boldsymbol{x}_{i(m)}\rho_{1}\left(y_{i}-\sum_{m=1}^{M}w_{m}\boldsymbol{x}_{i(m)}^{T}\Theta_{(m)}^{*}\right)\right\}\non\\
&&-\mathrm{E}\left\{\sum_{m=1}^{M}w_{m}\left(A_{(m)}^{-1}
V_{(m)}\right)^{T}\right\}\sum_{i=1}^{n}\boldsymbol{x}_{i(m)}\mathrm{E}\rho_{1}\left(y_{i}-\sum_{m=1}^{M}w_{m}\boldsymbol{x}_{i(m)}^{T}\Theta_{(m)}^{*}\right)\non\\
&=&\sum_{i=1}^{n}\mathrm{E}\rho\left(y_{i}-\sum_{m=1}^{M}w_{m}\boldsymbol{x}_{i(m)}^{T}\widehat{\Theta}_{(m)}\right)+\sum_{m=1}^{M}w_{m}\left[\mathrm{E}\left\{
V_{(m)}^{T}A_{(m)}^{-1}\widetilde{V}_{(m)}\right\}-\mathrm{E}\left\{V_{(m)}^{T}A_{(m)}^{-1}\right\}
\mathrm{E}\left(\widetilde{V}_{(m)}\right)\right]\non\\
&=&\sum_{i=1}^{n}\mathrm{E}\rho\left(y_{i}-\sum_{m=1}^{M}w_{m}\boldsymbol{x}_{i(m)}^{T}\widehat{\Theta}_{(m)}\right)\non\\
&&+\sum_{m=1}^{M}w_{m}
\mathrm{Cov}\left\{\sum_{i=1}^{n}\rho_{1}\left(y_{i}-\boldsymbol{x}_{i(m)}^{T}\Theta_{(m)}^{*}\right)\boldsymbol{x}_{i(m)}^{T}A_{(m)}^{-1},
\sum_{i=1}^{n}\rho_{1}\left(y_{i}-\sum_{m=1}^{M}w_{m}\boldsymbol{x}_{i(m)}^{T}\Theta_{(m)}^{*}\right)\boldsymbol{x}_{i(m)}\right\}\non\\
%&=&\sum_{i=1}^{n}\mathrm{E}\rho\left(y_{i}-\sum_{m=1}^{M}w_{m}\boldsymbol{x}_{i(m)}^{T}\widehat{\Theta}_{(m)}\right)\non\\
%&&+\sum_{m=1}^{M}w_{m}\mathrm{E}\left\{
%\sum_{i=1}^{n}\rho_{1}\left(y_{i}-\boldsymbol{x}_{i(m)}^{T}\Theta_{(m)}^{*}\right)\boldsymbol{x}_{i(m)}^{T}A_{(m)}^{-1}\cdot
%\sum_{i=1}^{n}\rho_{1}\left(y_{i}-\sum_{m=1}^{M}w_{m}\boldsymbol{x}_{i(m)}^{T}\Theta_{(m)}^{*}\right)\boldsymbol{x}_{i(m)}\right\}\non\\
&=&\sum_{i=1}^{n}\mathrm{E}\rho\left(y_{i}-\sum_{m=1}^{M}w_{m}\boldsymbol{x}_{i(m)}^{T}\widehat{\Theta}_{(m)}\right)\non\\
&&+\sum_{m=1}^{M}w_{m}\sum_{i=1}^{n}
\mathrm{Cov}\left\{\rho_{1}\left(y_{i}-\sum_{m=1}^{M}w_{m}\boldsymbol{x}_{i(m)}^{T}\Theta_{(m)}^{*}\right),
\rho_{1}\left(y_{i}-\boldsymbol{x}_{i(m)}^{T}\Theta_{(m)}^{*}\right)\right\}
\boldsymbol{x}_{i(m)}^{T}A_{(m)}^{-1}\boldsymbol{x}_{i(m)}.
\ee

Following \cite{Burman1995A},
$$\mathrm{Cov}\left\{\rho_{1}\left(y_{i}-\sum_{m=1}^{M}w_{m}\boldsymbol{x}_{i(m)}^{T}\Theta_{(m)}^{*}\right),
\rho_{1}\left(y_{i}-\boldsymbol{x}_{i(m)}^{T}\Theta_{(m)}^{*}\right)\right\}$$ can be approximated by the average
$$\frac{1}{n}\sum_{i=1}^{n}\mathrm{Cov}\left\{\rho_{1}\left(y_{i}-\sum_{m=1}^{M}w_{m}\boldsymbol{x}_{i(m)}^{T}\Theta_{(m)}^{*}\right),
\rho_{1}\left(y_{i}-\boldsymbol{x}_{i(m)}^{T}\Theta_{(m)}^{*}\right)\right\}$$
and
$A_{(m)}$ is replaced by
$$\left\{\frac{1}{n}\sum_{i=1}^{n}R_{2}\left(\mu_{i}-\boldsymbol{x}_{i(m)}^{T}\Theta_{(m)}^{*}\right)\right\}\sum_{i=1}^{n}\boldsymbol{x}_{i(m)}\boldsymbol{x}_{i(m)}^{T}.$$
Then
\be
&&\mathrm{E}\sum_{i=1}^{n}\rho\left(\widetilde{y}_{i}-\sum_{m=1}^{M}w_{m}\boldsymbol{x}_{i(m)}^{T}\widehat{\Theta}_{(m)}\right)\non\\
&\simeq&\sum_{i=1}^{n}\mathrm{E}\rho\left(y_{i}-\sum_{m=1}^{M}w_{m}\boldsymbol{x}_{i(m)}^{T}\widehat{\Theta}_{(m)}\right)\non\\
&&+\sum_{m=1}^{M}
\widetilde{C}_{\rho(m)}
w_{m}k_{m},
\ee
where $$\widetilde{C}_{\rho(m)}=\frac{\sum_{i=1}^{n}\mathrm{Cov}\left\{\rho_{1}\left(y_{i}-\sum_{m=1}^{M}w_{m}\boldsymbol{x}_{i(m)}^{T}\Theta_{(m)}^{*}\right),
\rho_{1}\left(y_{i}-\boldsymbol{x}_{i(m)}^{T}\Theta_{(m)}^{*}\right)\right\}}
{\sum_{i=1}^{n}R_{2}\left(\mu_{i}-\boldsymbol{x}_{i(m)}^{T}\Theta_{(m)}^{*}\right)}.$$
If we use $\boldsymbol{x}_{i}^{T}\Theta^{*}$ to approximate $\boldsymbol{x}_{i(m)}^{T}\Theta_{(m)}^{*}$ and $\sum_{m=1}^{M}w_{m}\boldsymbol{x}_{i(m)}^{T}\Theta_{(m)}^{*}$,
then
\be
\widetilde{C}_{\rho(m)}\simeq C_{\rho}
=\frac{\sum_{i=1}^{n}\mathrm{var}\left\{\rho_{1}\left(y_{i}-\boldsymbol{x}_{i}^{T}\Theta^{*}\right)\right\}}
{\sum_{i=1}^{n}R_{2}\left(\mu_{i}-\boldsymbol{x}_{i}^{T}\Theta^{*}\right)}.
\ee
We approximate $\sum_{i=1}^{n}\mathrm{E}\rho\left(y_{i}-\sum_{m=1}^{M}w_{m}\boldsymbol{x}_{i(m)}^{T}\Theta_{(m)}^{*}\right)$ by the sum itself.
In conclusion,
\be
&&\mathrm{E}\sum_{i=1}^{n}\rho\left(\widetilde{y}_{i}-\sum_{m=1}^{M}w_{m}\boldsymbol{x}_{i(m)}^{T}\widehat{\Theta}_{(m)}\right)\non\\
&\simeq&\sum_{i=1}^{n}\rho\left(y_{i}-\sum_{m=1}^{M}w_{m}\boldsymbol{x}_{i(m)}^{T}\widehat{\Theta}_{(m)}\right)\non\\
&&+C_{\rho}\sum_{m=1}^{M}
w_{m}k_{m}.
\ee
Therefore, we can use \eqref{C:fixed} to choose weight.
%%%%%%%%%%%%%%%%%%%%%%%%%%%%%%%%%%%%%%%%%%%%%%%%%%%%%%%%%%%%%%%%%%%%%%%%%%%%
\section{The derivation of \eqref{C:random}}\label{sec:B}
When $\boldsymbol{x}_{1},\ldots,\boldsymbol{x}_{n}$ are independently and identically distributed, we also have the conclusion \eqref{eq:rhom}-\eqref{eq:rhomhat}.
By the definition of $\Theta_{(m)}^{*}$ and Taylor expansion, for $\Theta_{(m)}$ near $\Theta_{(m)}^{*}$, we can derive that
\be\label{eqX:Erho}
&&\mathrm{E}\left\{\sum_{i=1}^{n}\rho\left(y_{i}-\sum_{m=1}^{M}w_{m}\boldsymbol{x}_{i(m)}^{T}\Theta_{(m)}\right)|\boldsymbol{x}_{i}\right\}\non\\
&=&\sum_{i=1}^{n}\mathrm{E}\left\{\rho\left(y_{i}-\sum_{m=1}^{M}w_{m}\boldsymbol{x}_{i(m)}^{T}\Theta_{(m)}^{*}\right)|\boldsymbol{x}_{i}\right\}\non\\
&&+\sum_{m=1}^{M}w_{m}\left(\Theta_{(m)}-\Theta_{(m)}^{*}\right)^{T}\sum_{i=1}^{n}\boldsymbol{x}_{i(m)}\mathrm{E}\left\{\rho_{1}\left(y_{i}-\sum_{m=1}^{M}w_{m}\boldsymbol{x}_{i(m)}^{T}\Theta_{(m)}^{*}\right)|\boldsymbol{x}_{i}\right\}\non\\
&&+\frac{1}{2}\left[\sum_{m=1}^{M}w_{m}\boldsymbol{x}_{i(m)}^{T}\left(\Theta_{(m)}-\Theta_{(m)}^{*}\right)\right]^{T}
R_{2}\left(\mu_{i}-\sum_{m=1}^{M}w_{m}\boldsymbol{x}_{i(m)}^{T}\Theta_{(m)}^{*}\right)\non\\
&&\left[\sum_{m=1}^{M}w_{m}\boldsymbol{x}_{i(m)}^{T}\left(\Theta_{(m)}-\Theta_{(m)}^{*}\right)\right]
\{1+o(1)\}.
\ee

Taking expectation over $\widetilde{y}$, the notation denoted by $\mathrm{\widetilde{E}}$ and from \eqref{eqX:Erho}, we obtain that
\be \label{eq:tildeErho}
&&\mathrm{\widetilde{E}}\sum_{i=1}^{n}\rho\left(\widetilde{y}_{i}-\sum_{m=1}^{M}w_{m}\boldsymbol{\widetilde{x}}_{i(m)}^{T}\widehat{\Theta}_{(m)}\right)\non\\
&=&\sum_{i=1}^{n}\mathrm{\widetilde{E}}\rho\left(\widetilde{y}_{i}-\sum_{m=1}^{M}w_{m}\boldsymbol{\widetilde{x}}_{i(m)}^{T}\Theta_{(m)}^{*}\right)\non\\
&&+\sum_{m=1}^{M}w_{m}\left(\widehat{\Theta}_{(m)}-\Theta_{(m)}^{*}\right)^{T}\sum_{i=1}^{n}\boldsymbol{\widetilde{x}}_{i(m)}\mathrm{\widetilde{E}}\rho_{1}\left(\widetilde{y}_{i}-\sum_{m=1}^{M}w_{m}\boldsymbol{\widetilde{x}}_{i(m)}^{T}\Theta_{(m)}^{*}\right)\non\\
&&+\frac{1}{2}\left[\sum_{m=1}^{M}w_{m}\boldsymbol{\widetilde{x}}_{i(m)}^{T}\left(\widehat{\Theta}_{(m)}-\Theta_{(m)}^{*}\right)\right]^{T}
R_{2}\left(\mu_{i}-\sum_{m=1}^{M}w_{m}\boldsymbol{\widetilde{x}}_{i(m)}^{T}\Theta_{(m)}^{*}\right)\non\\
&&\left[\sum_{m=1}^{M}w_{m}\boldsymbol{\widetilde{x}}_{i(m)}^{T}\left(\widehat{\Theta}_{(m)}-\Theta_{(m)}^{*}\right)\right]
\{1+o(1)\}.
\ee
Taking expectation for \eqref{eq:tildeErho},
\be\label{eq:tildewhat}
&&\mathrm{E}\sum_{i=1}^{n}\rho\left(\widetilde{y}_{i}-\sum_{m=1}^{M}w_{m}\boldsymbol{\widetilde{x}}_{i(m)}^{T}\widehat{\Theta}_{(m)}\right)\non\\
&=&\mathrm{E}\sum_{i=1}^{n}\rho\left(\widetilde{y}_{i}-\sum_{m=1}^{M}w_{m}\boldsymbol{\widetilde{x}}_{i(m)}^{T}\Theta_{(m)}^{*}\right)\non\\
&&+\mathrm{E}\left\{\sum_{m=1}^{M}w_{m}\left(-A_{(m)}^{-1}
V_{(m)}\right)^{T}\right\}\mathrm{E}\sum_{i=1}^{n}\boldsymbol{\widetilde{x}}_{i(m)}\rho_{1}\left(\widetilde{y}_{i}-\sum_{m=1}^{M}w_{m}\boldsymbol{\widetilde{x}}_{i(m)}^{T}\Theta_{(m)}^{*}\right)\non\\
&&+\frac{1}{2}\mathrm{E}\left[\left\{\sum_{m=1}^{M}w_{m}\boldsymbol{\widetilde{x}}_{i(m)}^{T}\left(-A_{(m)}^{-1}
V_{(m)}\right)\right\}^{T}
R_{2}\left(\mu_{i}-\sum_{m=1}^{M}w_{m}\boldsymbol{\widetilde{x}}_{i(m)}^{T}\Theta_{(m)}^{*}\right)\right.\non\\
&&\left.\left\{\sum_{m=1}^{M}w_{m}\boldsymbol{\widetilde{x}}_{i(m)}^{T}\left(-A_{(m)}^{-1}
V_{(m)}\right)\right\}\right]
\{1+o(1)\}.
\ee
Similar to the derivation of \eqref{eq:Erhomhat}, we have
\be\label{eq:what}
&&\mathrm{E}\sum_{i=1}^{n}\rho\left(y_{i}-\sum_{m=1}^{M}w_{m}\boldsymbol{x}_{i(m)}^{T}\widehat{\Theta}_{(m)}\right)\non\\
&=&\mathrm{E}\sum_{i=1}^{n}\rho\left(y_{i}-\sum_{m=1}^{M}w_{m}\boldsymbol{x}_{i(m)}^{T}\Theta_{(m)}^{*}\right)\non\\
&&+\mathrm{E}\left\{\sum_{m=1}^{M}w_{m}\left(-A_{(m)}^{-1}
V_{(m)}\right)^{T}\sum_{i=1}^{n}\boldsymbol{x}_{i(m)}\rho_{1}\left(y_{i}-\sum_{m=1}^{M}w_{m}\boldsymbol{x}_{i(m)}^{T}\Theta_{(m)}^{*}\right)\right\}\non\\
&&+\frac{1}{2}\mathrm{E}\left[\left\{\sum_{m=1}^{M}w_{m}\boldsymbol{x}_{i(m)}^{T}\left(-A_{(m)}^{-1}
V_{(m)}\right)\right\}^{T}
R_{2}\left(\mu_{i}-\sum_{m=1}^{M}w_{m}\boldsymbol{x}_{i(m)}^{T}\Theta_{(m)}^{*}\right)\right.\non\\
&&\left.\left\{\sum_{m=1}^{M}w_{m}\boldsymbol{x}_{i(m)}^{T}\left(-A_{(m)}^{-1}
V_{(m)}\right)\right\}\right]
\{1+o(1)\}+o_{p}(1).
\ee
Combining \eqref{eq:tildewhat} and \eqref{eq:what}, we have
\be
&&\mathrm{E}\sum_{i=1}^{n}\rho\left(\widetilde{y}_{i}-\sum_{m=1}^{M}w_{m}\boldsymbol{\widetilde{x}}_{i(m)}^{T}\widehat{\Theta}_{(m)}\right)\non\\
&\simeq&\sum_{i=1}^{n}\mathrm{E}\rho\left(y_{i}-\sum_{m=1}^{M}w_{m}\boldsymbol{x}_{i(m)}^{T}\widehat{\Theta}_{(m)}\right)\non\\
&&+\mathrm{E}\left\{\sum_{m=1}^{M}w_{m}\left(A_{(m)}^{-1}
V_{(m)}\right)^{T}\sum_{i=1}^{n}\boldsymbol{x}_{i(m)}\rho_{1}\left(y_{i}-\sum_{m=1}^{M}w_{m}\boldsymbol{x}_{i(m)}^{T}\Theta_{(m)}^{*}\right)\right\}\non\\
&&-\mathrm{E}\left\{\sum_{m=1}^{M}w_{m}\left(A_{(m)}^{-1}
V_{(m)}\right)^{T}\right\}\mathrm{E}\sum_{i=1}^{n}\boldsymbol{\widetilde{x}}_{i(m)}\rho_{1}\left(\widetilde{y}_{i}-\sum_{m=1}^{M}w_{m}\boldsymbol{\widetilde{x}}_{i(m)}^{T}\Theta_{(m)}^{*}\right)\non\\
&=&\sum_{i=1}^{n}\mathrm{E}\rho\left(y_{i}-\sum_{m=1}^{M}w_{m}\boldsymbol{x}_{i(m)}^{T}\widehat{\Theta}_{(m)}\right)\non\\
&&+\mathrm{E}\left\{\sum_{m=1}^{M}w_{m}\left(A_{(m)}^{-1}
V_{(m)}\right)^{T}\sum_{i=1}^{n}\boldsymbol{x}_{i(m)}\rho_{1}\left(y_{i}-\sum_{m=1}^{M}w_{m}\boldsymbol{x}_{i(m)}^{T}\Theta_{(m)}^{*}\right)\right\}\non\\
&&-\mathrm{E}\left\{\sum_{m=1}^{M}w_{m}\left(A_{(m)}^{-1}
V_{(m)}\right)^{T}\right\}\mathrm{E}\sum_{i=1}^{n}\boldsymbol{x}_{i(m)}\rho_{1}\left(y_{i}-\sum_{m=1}^{M}w_{m}\boldsymbol{x}_{i(m)}^{T}\Theta_{(m)}^{*}\right)\non\\
&=&\sum_{i=1}^{n}\mathrm{E}\rho\left(y_{i}-\sum_{m=1}^{M}w_{m}\boldsymbol{x}_{i(m)}^{T}\widehat{\Theta}_{(m)}\right)\non\\
&&+\sum_{m=1}^{M}w_{m}\mathrm{tr}\left[
\mathrm{Cov}\left\{\sum_{i=1}^{n}\rho_{1}\left(y_{i}-\boldsymbol{x}_{i(m)}^{T}\Theta_{(m)}^{*}\right)A_{(m)}^{-1}\boldsymbol{x}_{i(m)},
\sum_{i=1}^{n}\boldsymbol{x}_{i(m)}\rho_{1}\left(y_{i}-\sum_{m=1}^{M}w_{m}\boldsymbol{x}_{i(m)}^{T}\Theta_{(m)}^{*}\right)\right\}\right]\non\\
&=&\sum_{i=1}^{n}\mathrm{E}\rho\left(y_{i}-\sum_{m=1}^{M}w_{m}\boldsymbol{x}_{i(m)}^{T}\widehat{\Theta}_{(m)}\right)\non\\
&&+\sum_{m=1}^{M}w_{m}\sum_{i=1}^{n}\mathrm{tr}\left[
\mathrm{Cov}\left\{\rho_{1}\left(y_{i}-\boldsymbol{x}_{i(m)}^{T}\Theta_{(m)}^{*}\right)A_{(m)}^{-1}\boldsymbol{x}_{i(m)},
\boldsymbol{x}_{i(m)}\rho_{1}\left(y_{i}-\sum_{m=1}^{M}w_{m}\boldsymbol{x}_{i(m)}^{T}\Theta_{(m)}^{*}\right)\right\}\right]\non\\
&=&\sum_{i=1}^{n}\mathrm{E}\rho\left(y_{i}-\sum_{m=1}^{M}w_{m}\boldsymbol{x}_{i(m)}^{T}\widehat{\Theta}_{(m)}\right)\non\\
&&+\sum_{m=1}^{M}w_{m}\sum_{i=1}^{n}\mathrm{E}\left\{\rho_{1}\left(y_{i}-\boldsymbol{x}_{i(m)}^{T}\Theta_{(m)}^{*}\right)\rho_{1}\left(y_{i}-\sum_{m=1}^{M}w_{m}\boldsymbol{x}_{i(m)}^{T}\Theta_{(m)}^{*}\right)
\boldsymbol{x}_{i(m)}^{T}A_{(m)}^{-1}\boldsymbol{x}_{i(m)}\right\}\non\\
&&-\mathrm{E}\left\{\rho_{1}\left(y_{i}-\boldsymbol{x}_{i(m)}^{T}\Theta_{(m)}^{*}\right)\boldsymbol{x}_{i(m)}^{T}A_{(m)}^{-1}\right\}\mathrm{E}\left\{\boldsymbol{x}_{i(m)}\rho_{1}\left(y_{i}-\sum_{m=1}^{M}w_{m}\boldsymbol{x}_{i(m)}^{T}\Theta_{(m)}^{*}\right)\right\}\non\\
&=&\sum_{i=1}^{n}\mathrm{E}\rho\left(y_{i}-\sum_{m=1}^{M}w_{m}\boldsymbol{x}_{i(m)}^{T}\widehat{\Theta}_{(m)}\right)\non\\
&&+\sum_{m=1}^{M}w_{m}\sum_{i=1}^{n}\mathrm{E}\left\{\rho_{1}\left(y_{i}-\boldsymbol{x}_{i(m)}^{T}\Theta_{(m)}^{*}\right)\rho_{1}\left(y_{i}-\sum_{m=1}^{M}w_{m}\boldsymbol{x}_{i(m)}^{T}\Theta_{(m)}^{*}\right)
\boldsymbol{x}_{i(m)}^{T}A_{(m)}^{-1}\boldsymbol{x}_{i(m)}\right\}\non\\
\ee

Following \cite{Burman1995A},
$A_{(m)}$ is replaced by
$$\left\{\frac{1}{n}\sum_{i=1}^{n}R_{2}\left(\mu_{i}-\boldsymbol{x}_{i(m)}^{T}\Theta_{(m)}^{*}\right)\right\}\sum_{i=1}^{n}\boldsymbol{x}_{i(m)}\boldsymbol{x}_{i(m)}^{T}.$$
Then
\be
&&\sum_{m=1}^{M}w_{m}\sum_{i=1}^{n}\mathrm{E}\left\{\rho_{1}\left(y_{i}-\boldsymbol{x}_{i(m)}^{T}\Theta_{(m)}^{*}\right)\rho_{1}\left(y_{i}-\sum_{m=1}^{M}w_{m}\boldsymbol{x}_{i(m)}^{T}\Theta_{(m)}^{*}\right)
\boldsymbol{x}_{i(m)}^{T}A_{(m)}^{-1}\boldsymbol{x}_{i(m)}\right\}\non\\
&\approx& \sum_{m=1}^{M}w_{m}\sum_{i=1}^{n}\mathrm{E}\left\{\rho_{1}\left(y_{i}-\boldsymbol{x}_{i(m)}^{T}\Theta_{(m)}^{*}\right)\rho_{1}\left(y_{i}-\sum_{m=1}^{M}w_{m}\boldsymbol{x}_{i(m)}^{T}\Theta_{(m)}^{*}\right)
\right.\non\\
&&\left.\left\{\frac{1}{n}\sum_{i=1}^{n}R_{2}\left(\mu_{i}-\boldsymbol{x}_{i(m)}^{T}\Theta_{(m)}^{*}\right)\right\}^{-1}\boldsymbol{x}_{i(m)}^{T}\left(\sum_{i=1}^{n}\boldsymbol{x}_{i(m)}\boldsymbol{x}_{i(m)}^{T}\right)^{-1}\boldsymbol{x}_{i(m)}\right\}.
\ee
Next,
$$\boldsymbol{x}_{i(m)}^{T}\left(\sum_{i=1}^{n}\boldsymbol{x}_{i(m)}\boldsymbol{x}_{i(m)}^{T}\right)^{-1}\boldsymbol{x}_{i(m)}$$ can be approximated by the average
$$\frac{1}{n}\sum_{i=1}^{n}\boldsymbol{x}_{i(m)}^{T}\left(\sum_{i=1}^{n}\boldsymbol{x}_{i(m)}\boldsymbol{x}_{i(m)}^{T}\right)^{-1}\boldsymbol{x}_{i(m)}=\frac{k_{m}}{n}.$$
Therefore,
\be
&&\sum_{m=1}^{M}w_{m}\sum_{i=1}^{n}\mathrm{E}\left\{\rho_{1}\left(y_{i}-\boldsymbol{x}_{i(m)}^{T}\Theta_{(m)}^{*}\right)\rho_{1}\left(y_{i}-\sum_{m=1}^{M}w_{m}\boldsymbol{x}_{i(m)}^{T}\Theta_{(m)}^{*}\right)
\boldsymbol{x}_{i(m)}^{T}A_{(m)}^{-1}\boldsymbol{x}_{i(m)}\right\}\non\\
&\approx& \sum_{m=1}^{M}w_{m}k_{m}\mathrm{E}\left\{\rho_{1}\left(y_{i}-\boldsymbol{x}_{i(m)}^{T}\Theta_{(m)}^{*}\right)\rho_{1}\left(y_{i}-\sum_{m=1}^{M}w_{m}\boldsymbol{x}_{i(m)}^{T}\Theta_{(m)}^{*}\right)
\left\{\frac{1}{n}\sum_{i=1}^{n}R_{2}\left(\mu_{i}-\boldsymbol{x}_{i(m)}^{T}\Theta_{(m)}^{*}\right)\right\}^{-1}\right\}\non\\
&\doteq& \sum_{m=1}^{M}w_{m}k_{m}C_{\rho(m)}.
\ee
%Let $\widehat{\varepsilon}_{i(m)}=y_{i}-\boldsymbol{x}_{i(m)}^{T}\widehat{\Theta}_{(m)}$.
%The term $C_{\rho(m)}$
%can be estimated by
%$$\widehat{C}_{\rho(m)}=\frac{1}{n}\sum_{i=1}^{n}\left\{\rho_{1}\left(\widehat{\varepsilon}_{i(m)}\right)\rho_{1}\left(\sum_{m=1}^{M}w_{m}\widehat{\varepsilon}_{i(m)}\right)
%\left\{\frac{1}{n}\sum_{i=1}^{n}\widehat{R}_{2}\left(\widehat{\varepsilon}_{i(m)}\right)\right\}^{-1}\right\},$$
%where
%$\widehat{R}_{2}$ is an approximation for $R_{2}$.
Therefore, \eqref{C:random} can be applied to choose weight.

\bibliographystyle{Chicago}
\bibliography{RCp}

\begin{thebibliography}{}

\bibitem[\protect\citeauthoryear{Agostinelli}{Agostinelli}{2002a}]{Agostinelli2002Robust}
Agostinelli, C. (2002a).
\newblock Robust model selection in regression via weighted likelihood
  methodology.
\newblock {\em Statistics and Probability Letters\/}~{\em 56\/}(3), 289--300.

\bibitem[\protect\citeauthoryear{Agostinelli}{Agostinelli}{2002b}]{Agostinelli2002Robuststep}
Agostinelli, C. (2002b).
\newblock Robust stepwise regression.
\newblock {\em Journal of Applied Statistics\/}~{\em 29\/}(6), 825--840.

\bibitem[\protect\citeauthoryear{Agostinelli and Markatou}{Agostinelli and
  Markatou}{1998}]{Agostinelli1998A}
Agostinelli, C. and M.~Markatou (1998).
\newblock A one-step robust estimator for regression based on the weighted
  likelihood reweighting scheme.
\newblock {\em Statistics and Probability Letters\/}~{\em 37\/}(4), 341--350.

\bibitem[\protect\citeauthoryear{Akaike}{Akaike}{1973}]{Akaike1973Information}
Akaike, H. (1973).
\newblock Information theory and an extension of the maximum likelihood
  principle.
\newblock In {\em International Symposium on Information Theory}, pp.\
  267--281.

\bibitem[\protect\citeauthoryear{Ando and Li}{Ando and
  Li}{2014}]{ando2014model}
Ando, T. and K.-C. Li (2014).
\newblock A model-averaging approach for high-dimensional regression.
\newblock {\em Journal of the American Statistical Association\/}~{\em
  109\/}(505), 254--265.

\bibitem[\protect\citeauthoryear{Andrews}{Andrews}{1974}]{Andrews1974A}
Andrews, D.~F. (1974).
\newblock A robust method for multiple linear regression.
\newblock {\em Technometrics\/}~{\em 16\/}(4), 523--531.

\bibitem[\protect\citeauthoryear{Buckland, Burnham, and Augustin}{Buckland
  et~al.}{1997}]{Buckland1997Model}
Buckland, S.~T., K.~P. Burnham, and N.~H. Augustin (1997).
\newblock Model selection: An integral part of inference.
\newblock {\em Biometrics\/}~{\em 53\/}(2), 603--618.

\bibitem[\protect\citeauthoryear{Burman and Nolan}{Burman and
  Nolan}{1995}]{Burman1995A}
Burman, P. and D.~Nolan (1995).
\newblock A general akaike-type criterion for model selection in robust
  regression.
\newblock {\em Biometrika\/}~{\em 82\/}(4), 877--886.

\bibitem[\protect\citeauthoryear{Claeskens, Croux, and Kerckhoven}{Claeskens
  et~al.}{2006}]{claeskens2006logit}
Claeskens, G., C.~Croux, and J.~Kerckhoven (2006).
\newblock Variable selection for logistic regression using a prediction-focused
  information criterion.
\newblock {\em Biometrics\/}~{\em 62\/}(4), 972--979.

\bibitem[\protect\citeauthoryear{Hampel}{Hampel}{1983}]{Hampel1983Some}
Hampel, F.~R. (1983).
\newblock Some aspects of model choice in robust statistics.
\newblock In {\em Proceedings of the 44th Session of the ISI, Madrid, Book 2},
  pp.\  767--771.

\bibitem[\protect\citeauthoryear{Hampel, Ronchetti, Rousseeuw, and
  Stahel}{Hampel et~al.}{1986}]{hampel1986robust}
Hampel, F.~R., E.~M. Ronchetti, P.~J. Rousseeuw, and W.~A. Stahel (1986).
\newblock {\em {Robust Statistics: The Approach Based on Influence Functions}}.
\newblock New York: John Wiley.

\bibitem[\protect\citeauthoryear{Hansen}{Hansen}{2007}]{hansen2007least}
Hansen, B.~E. (2007).
\newblock Least squares model averaging.
\newblock {\em Econometrica\/}~{\em 75\/}(4), 1175--1189.

\bibitem[\protect\citeauthoryear{Hansen}{Hansen}{2008}]{hansen2008joe}
Hansen, B.~E. (2008).
\newblock Least squares forecast averaging.
\newblock {\em Journal of Econometrics\/}~{\em 146}, 342--350.

\bibitem[\protect\citeauthoryear{Hansen and Racine}{Hansen and
  Racine}{2012}]{hansen2012jackknife}
Hansen, B.~E. and J.~S. Racine (2012).
\newblock Jackknife model averaging.
\newblock {\em Journal of Econometrics\/}~{\em 167\/}(1), 38--46.

\bibitem[\protect\citeauthoryear{Hawkins}{Hawkins}{1980}]{Hawkins1980}
Hawkins, D.~M. (1980).
\newblock {\em {Identification of Outliers}}.
\newblock Springer Science and Business Media, B.V.

\bibitem[\protect\citeauthoryear{Hawkins, Dan, and Kass}{Hawkins
  et~al.}{1984}]{Hawkins1984Location}
Hawkins, D.~M., B.~Dan, and G.~V. Kass (1984).
\newblock Location of several outliers in multiple-regression data using
  elemental sets.
\newblock {\em Technometrics\/}~{\em 26\/}(3), 197--208.

\bibitem[\protect\citeauthoryear{Hjort and Claeskens}{Hjort and
  Claeskens}{2003}]{hjort2003averaging}
Hjort, N.~L. and G.~Claeskens (2003).
\newblock Frequentist model average estimators.
\newblock {\em Journal of the American Statistical Association\/}~{\em
  98\/}(464), 879--899.

\bibitem[\protect\citeauthoryear{Hjort and Claeskens}{Hjort and
  Claeskens}{2006}]{hjort2006focused}
Hjort, N.~L. and G.~Claeskens (2006).
\newblock Focused information criteria and model averaging for the {C}ox hazard
  regression model.
\newblock {\em Journal of the American Statistical Association\/}~{\em
  101\/}(476), 1449--1464.

\bibitem[\protect\citeauthoryear{Huber}{Huber}{1964}]{Huber1964Robust}
Huber, P.~J. (1964).
\newblock Robust estimation of a location parameter.
\newblock {\em Annals of Mathematical Statistics\/}~{\em 35\/}(1), 73--101.

\bibitem[\protect\citeauthoryear{Liang, Zou, Wan, and Zhang}{Liang
  et~al.}{2011}]{liang2011optimal}
Liang, H., G.~H. Zou, A.~T.~K. Wan, and X.~Y. Zhang (2011).
\newblock Optimal weight choice for frequentist model average estimators.
\newblock {\em Journal of the American Statistical Association\/}~{\em
  106\/}(495), 1053--1066.

\bibitem[\protect\citeauthoryear{Lindsay}{Lindsay}{1994}]{lindsay1994}
Lindsay, B.~G. (1994).
\newblock Efficiency versus robustness: The case for minimum hellinger distance
  and related methods.
\newblock {\em Annals of Statistics\/}~{\em 22\/}(2), 1081--1114.

\bibitem[\protect\citeauthoryear{Lu and Su}{Lu and Su}{2015}]{lu2015jackknife}
Lu, X. and L.~Su (2015).
\newblock Jackknife model averaging for quantile regressions.
\newblock {\em Journal of Econometrics\/}~{\em 188\/}(1), 40--58.

\bibitem[\protect\citeauthoryear{Mallows}{Mallows}{1973}]{Mallows1973Some}
Mallows, C.~L. (1973).
\newblock Some comments on {$C_{p}$}.
\newblock {\em Technometrics\/}~{\em 15\/}(4), 661--675.

\bibitem[\protect\citeauthoryear{M\"{u}ller and Welsh}{M\"{u}ller and
  Welsh}{2005}]{M2005Outlier}
M\"{u}ller, S. and A.~H. Welsh (2005).
\newblock Outlier robust model selection in linear regression.
\newblock {\em Journal of the American Statistical Association\/}~{\em
  100\/}(472), 1297--1310.

\bibitem[\protect\citeauthoryear{Ronchetti}{Ronchetti}{1985}]{Ronchetti1985Robust}
Ronchetti, E. (1985).
\newblock Robust model selection in regression.
\newblock {\em Statistics and Probability Letters\/}~{\em 3\/}(1), 21--23.

\bibitem[\protect\citeauthoryear{Ronchetti}{Ronchetti}{1997}]{Ronchetti1997Robustness}
Ronchetti, E. (1997).
\newblock Robustness aspects of model choice.
\newblock {\em Statistica Sinica\/}~{\em 7\/}(2), 327--338.

\bibitem[\protect\citeauthoryear{Ronchetti, Field, and Blanchard}{Ronchetti
  et~al.}{1997}]{Ronchetti1997Robust}
Ronchetti, E., C.~Field, and W.~Blanchard (1997).
\newblock Robust linear model selection by cross-validation.
\newblock {\em Journal of the American Statistical Association\/}~{\em
  92\/}(439), 1017--1023.

\bibitem[\protect\citeauthoryear{Ronchetti and Staudte}{Ronchetti and
  Staudte}{1994}]{Ronchetti1994A}
Ronchetti, E. and R.~G. Staudte (1994).
\newblock A robust version of {Mallow's C$_{p}$}.
\newblock {\em Journal of the American Statistical Association\/}~{\em
  89\/}(426), 550--559.

\bibitem[\protect\citeauthoryear{Schwarz}{Schwarz}{1978}]{Schwarz1978The}
Schwarz, G. (1978).
\newblock Estimating the dimension of a model.
\newblock {\em Annals of Statistics\/}~{\em 6\/}(2), 461--464.

\bibitem[\protect\citeauthoryear{Shao}{Shao}{1993}]{Shao1993Linear}
Shao, J. (1993).
\newblock Linear model selection by cross-validation.
\newblock {\em Journal of the American Statistical Association\/}~{\em
  88\/}(422), 486--494.

\bibitem[\protect\citeauthoryear{Sommer and Huggins}{Sommer and
  Huggins}{1996}]{Sommer1996Variables}
Sommer, S. and R.~M. Huggins (1996).
\newblock Variables selection using the wald test and a robust {C$_{p}$}.
\newblock {\em Journal of the Royal Statistical Society. Series C (Applied
  Statistics)\/}~{\em 45\/}(1), 15--29.

\bibitem[\protect\citeauthoryear{Sommer and Staudte}{Sommer and
  Staudte}{1995}]{Sommer1995Robust}
Sommer, S. and R.~G. Staudte (1995).
\newblock Robust variable selection in regression in the presence of outliers
  and leverage points.
\newblock {\em Australian Journal of Statistics\/}~{\em 37\/}(3), 323--336.

\bibitem[\protect\citeauthoryear{Stone}{Stone}{1974}]{Stone1974Cross}
Stone, M. (1974).
\newblock Cross-validatory choice and assessment of statistical predictions.
\newblock {\em Journal of the Royal Statistical Society, Series B (Statistical
  Methodology)\/}~{\em 36\/}(2), 111--147.

\bibitem[\protect\citeauthoryear{Wald}{Wald}{1944}]{Wald1944On}
Wald, A. (1944).
\newblock On a statistical problem arising in the classification of an
  individual into one of two groups.
\newblock {\em Annals of Mathematical Statistics\/}~{\em 15\/}(2), 145--162.

\bibitem[\protect\citeauthoryear{Wan, Zhang, and Wang}{Wan
  et~al.}{2014}]{wan2014ijf}
Wan, A. T.~K., X.~Y. Zhang, and S.~Wang (2014).
\newblock Frequentist model averaging for multinomial and ordered logit models.
\newblock {\em International Journal of Forecasting\/}~{\em 30\/}(1), 118--128.

\bibitem[\protect\citeauthoryear{Wan, Zhang, and Zou}{Wan
  et~al.}{2010}]{wan2010least}
Wan, A. T.~K., X.~Y. Zhang, and G.~H. Zou (2010).
\newblock Least squares model averaging by mallows criterion.
\newblock {\em Journal of Econometrics\/}~{\em 156\/}(2), 277--283.

\bibitem[\protect\citeauthoryear{Wisnowski, Simpson, Montgomery, and
  Runger}{Wisnowski et~al.}{2003}]{Wisnowski2003Resampling}
Wisnowski, J.~W., J.~R. Simpson, D.~C. Montgomery, and G.~C. Runger (2003).
\newblock Resampling methods for variable selection in robust regression.
\newblock {\em Computational Statistics and Data Analysis\/}~{\em 43\/}(3),
  341--355.

\bibitem[\protect\citeauthoryear{Yang}{Yang}{2001}]{yang2001adaptive}
Yang, Y.~H. (2001).
\newblock Adaptive regression by mixing.
\newblock {\em Journal of the American Statistical Association\/}~{\em
  96\/}(454), 574--588.

\bibitem[\protect\citeauthoryear{Zhang, L., Zou, and Liang}{Zhang
  et~al.}{2016}]{zhang2016optimal}
Zhang, X.~Y., Y.~D. L., G.~H. Zou, and H.~Liang (2016).
\newblock Optimal model averaging estimation for generalized linear models and
  generalized linear mixed-effects model.
\newblock {\em Journal of the American Statistical Association\/}~{\em
  111\/}(516), 1775--1790.

\bibitem[\protect\citeauthoryear{Zhang and Liang}{Zhang and
  Liang}{2011}]{zhang2011focused}
Zhang, X.~Y. and H.~Liang (2011).
\newblock Focused information criterion and model averaging for generalized
  additive partial linear models.
\newblock {\em Annals of Statistics\/}~{\em 39\/}(1), 174--200.

\bibitem[\protect\citeauthoryear{Zhang, Wan, and Zhou}{Zhang
  et~al.}{2012}]{zhang2012focused}
Zhang, X.~Y., A.~T.~K. Wan, and S.~Z. Zhou (2012).
\newblock Focused information criteria, model selection and model averaging in
  a {T}obit model with a non-zero threshold.
\newblock {\em Journal of Business and Economic Statistics\/}~{\em 30\/}(1),
  132--142.

\bibitem[\protect\citeauthoryear{Zhang, Wan, and Zou}{Zhang
  et~al.}{2013}]{zhang2013jackknife}
Zhang, X.~Y., A.~T.~K. Wan, and G.~H. Zou (2013).
\newblock Model averaging by jackknife criterion in models with dependent data.
\newblock {\em Journal of Econometrics\/}~{\em 174\/}(2), 82--94.

\bibitem[\protect\citeauthoryear{Zhang, Zou, and Carroll}{Zhang
  et~al.}{2015}]{zhang2015kullback}
Zhang, X.~Y., G.~H. Zou, and R.~Carroll (2015).
\newblock Model averaging based on {Kullback-Leibler} distance.
\newblock {\em Statistica Sinica\/}~{\em 25\/}(4), 1583--1598.

\bibitem[\protect\citeauthoryear{Zhang, Zou, and Liang}{Zhang
  et~al.}{2014}]{zhang2014mixed}
Zhang, X.~Y., G.~H. Zou, and H.~Liang (2014).
\newblock Model averaging and weight choice in linear mixed effects models.
\newblock {\em Biometrika\/}~{\em 101\/}(1), 205--218.

\end{thebibliography}
\end{document}